\documentclass[a4paper,fleqn,usenatbib]{mnras}

\usepackage{newtxtext,newtxmath}
\usepackage[T1]{fontenc}
\usepackage{ae,aecompl}

\usepackage{graphicx}	% Including figure files
\usepackage{amsmath}	% Advanced maths commands
\title[Dust porosity in galaxies]{Evolution of dust grain size distribution and grain porosity in galaxies}

\author[H. Hirashita \& V. B. Il'in]{
Hiroyuki Hirashita$^1$\thanks{E-mail: hirashita@asiaa.sinica.edu.tw} and
Vladimir B. Il'in$^{2,3,4}$
\\
$^{1}$Institute of Astronomy and Astrophysics, Academia Sinica,
Astronomy--Mathematics Building, No.\ 1, Section 4,
Roosevelt Road, Taipei 10617, Taiwan\\
$^{2}$St Petersburg State University, Universitetskij Pr. 28, St Petersburg 198504, Russia\\
$^3$Pulkovo Observatory, Pulkovskoe Sh. 65/1, St Petersburg 196140, Russia\\
$^4$St Petersburg University of Aerospace Instrumentation, Bol. Morskaya 67,
St Petersburg 190000, Russia\\
}

\date{Accepted XXX. Received YYY; in original form ZZZ}

\pubyear{2021}

\begin{document}
\label{firstpage}
\pagerange{\pageref{firstpage}--\pageref{lastpage}}
\maketitle

\begin{abstract}
The radiative properties of interstellar dust are affected not only by the
grain size distribution but also by the grain porosity.
%%The evolution of these two quantities (grain size distribution and porosity) is
%%mutually related through interstellar processing.
We develop a model for the evolution
of size-dependent grain porosity and grain size distribution over the entire history of galaxy evolution.
We include stellar dust production, supernova dust destruction, shattering,
coagulation, and accretion. Coagulation is {assumed to be} the source of grain porosity.
We {use a one-zone model} with a constant dense gas fraction
($\eta_\mathrm{dense}$), which regulates the balance between shattering and coagulation. 
We find that porosity develops after small grains are sufficiently created by
the interplay between shattering and accretion (at age $t\sim 1$~Gyr
for star formation time-scale $\tau_\mathrm{SF}=5$~Gyr) and are coagulated.
The filling factor drops down to 0.3
at grain radii $\sim 0.03~\micron$ for $\eta_\mathrm{dense}=0.5$.
The grains are more porous for smaller $\eta_\mathrm{dense}$ because
{small grains, from which porous coagulated grains form, are more abundant.}
We also calculate the extinction curves based on the above results.
The porosity steepens the extinction curve significantly for silicate, but not much for
amorphous carbon.
The porosity also increases the collisional cross-sections and produces slightly more large
grains through the enhanced coagulation; however, the extinction curve does not
necessarily become flatter because of the steepening effect by porosity.
We also discuss the implication of our results for
the Milky Way extinction curve.
\end{abstract}


%%%%%%%%%%%%%%%%% BODY OF PAPER %%%%%%%%%%%%%%%%%%
\section{Introduction}\label{sec:intro}

Dust grains play an important role in various processes in the
interstellar medium (ISM) of galaxies.
As a galaxy is enriched with metals, the dust abundance becomes higher,
and the effects of dust on the following processes become more prominent.
Dust acts as a catalyst for the formation of various molecular species, especially
H$_2$
\citep[e.g.][]{Gould:1963aa,Cazaux:2004aa}.
Dust also strongly modifies the spectral energy distributions (SEDs) of galaxies
through the extinction of ultraviolet (UV)--optical stellar light and the reemission at
infrared (IR) wavelengths \citep[e.g.][]{Desert:1990aa,Takeuchi:2005aa}.
Radiation pressure on dust grains could affect galaxy-scale gas dynamics through
gas--grain coupling \citep[e.g.][]{Ferrara:1991aa,Murray:2011aa}.
The above processes are governed by the total surface area
\cite[e.g.][]{Yamasawa:2011aa,Chen:2018aa} and/or the extinction
cross-section \citep[e.g.][]{Draine:1984aa}, both of which are determined by the
grain size distribution. Thus, even with the same total dust abundance, the efficiencies of
the above-mentioned processes change with the grain size distribution.

The dust evolution is governed not only by the stellar dust production but also
by various processes in the ISM \citep[e.g.][]{Lisenfeld:1998aa,Draine:2009ab}.
Among the interstellar processing mechanisms, dust growth by the accretion of gas-phase metals
in the dense ISM is regarded as the dominant source for the
dust abundance in evolved (metal-enriched) galaxies
\citep[e.g.][]{Draine:1990aa,Dwek:1998aa,Hirashita:1999aa,Zhukovska:2008aa,Inoue:2011aa,Mattsson:2014aa}.
This process has also been
verified by some experiments \citep{Rouille:2014aa,Fulvio:2017aa,Rouille:2020aa}.
Dust growth by accretion is also an important mechanism in explaining the interstellar metal depletion
\citep{Weingartner:1999aa}.
The decrease of the dust abundance (dust-to-gas ratio) is, on the other hand, caused
by  dust destruction in supernova (SN) shocks \citep[e.g.][]{McKee:1989aa}.
The grain size distribution is further modified by coagulation in the dense ISM and
shattering in the diffuse ISM \citep[e.g.][]{ODonnell:1997aa,Hirashita:2009ab}.
These two processes determine the relative abundance between
large and small grains \citep{Hirashita:2009ab}. After a pioneering work by
\citet{ODonnell:1997aa},
a theoretical framework that includes all the above processes to calculate the
evolution of grain size distribution has been constructed by 
\citet{Asano:2013aa}, \citet{Nozawa:2015aa}, and \citet[][hereafter HA19]{Hirashita:2019aa}.

In addition to the grain size distribution, inhomogeneity within individual grains
in terms of the composition and the shape also affects the observed properties of
dust grains such as extinction curves (wavelength dependence of dust extinction).
\citet{Mathis:1989aa} assumed interstellar grains
to be aggregates of multiple species including vacuum (or voids), and explained the dust properties
in the Milky Way (MW).
Some studies also suggested
fluffy large grains based on the observations of the X-ray scattering halos around
point sources,
such as X-ray binaries \citep{Woo:1994aa} and Nova Cygni 1992 \citep{Mathis:1995aa},
although compact grains are also shown to be consistent with the
X-ray halo strengths based on further detailed calculations of
grain optical properties \citep{Smith:1998aa,Draine:2003aa}. In these interpretations,
there is an uncertainty arising from the distribution of dust
in the line of sight \citep{Draine:2003ab}.
On the other hand, from the theoretical point of view,
\citet[][hereafter H21]{Hirashita:2021aa} argued that coagulation develops fluffy or porous
grains in the ISM as already shown in the calculations of proto-planetary discs
\citep[e.g.][]{Okuzumi:2009aa,Okuzumi:2012aa} and of dense molecular cloud cores
\citep{Ormel:2009aa}.
Fluffy or porous grains produced in these dense environments
predict different extinction curves from compact (non-porous) grains
\citep{Ormel:2011aa,Kataoka:2014aa,Tazaki:2016aa,Lefevre:2020aa}.

Studying porous grains in the ISM was motivated by the limited available metals
(especially carbon). \citet{Mathis:1996aa}  found that, if 25--65 per cent of vacuum
is included in the dust grains, the least solid materials
(heavy elements) are required to explain the MW extinction curve.
%%Very fluffy grains are excluded because their extinction (absorption + scattering)
%%per unit dust mass is too low.
However,
\citet{Dwek:1997aa} argued the importance of fitting the IR dust emission SED
simultaneously, and showed that the above model of
\citet{Mathis:1996aa} overpredicts the optical--UV extinction.
\citet{Li:2005aa} more robustly showed using the Kramers--Kronig relations
that, if we use the Galactic metal abundance
derived from B stars, the observed extinction is still underproduced.
However, updated
extinction-to-column density ratios and protosolar metal abundances have
allowed us to construct a dust model consistent with
the metal abundance constraints \citep{Draine:2021ab,Hensley:2021aa,Zuo:2021aa}.
We still emphasize that the above studies did not necessarily give a definite
understanding of the grain porosity; for example, they addressed
neither \textit{grain-size-dependent} porosity nor connection with dust evolution.

The porosity of interstellar dust grains should be determined as a consequence of dust evolution
and should be dependent on the grain size.
Our previous study mentioned above (H21) solved the evolution of grain size distribution
and porosity by taking shattering and coagulation into account. Using H21's framework,
we are able to predict not only the porosity but also its grain size dependence.
Thus, in H21's framework, the porosity is not a
free parameter any more but a quantity predicted {along} with
the grain size distribution. Such a {comprehensive} understanding is particularly important
because the observed optical properties of dust depend on both
grain size distribution and grain porosity.
H21 also emphasized that the interplay between shattering and coagulation is important
in creating porous grains because small grains continuously supplied by shattering
become the ingredients for fluffy or porous grains formed by coagulation.
This means that including multiple processes for dust evolution is essential in understanding
the evolution of grain porosity.

{It is yet to be clarified if the grain porosity created by the interplay between coagulation
and shattering is maintained in the presence of other processes.}
H21 focused on shattering and coagulation, but did not include dust enrichment
and destruction that occur as a result of galaxy evolution (i.e.\ star formation
and metal enrichment). Specifically, among the
processes mentioned above, stellar dust production, dust growth by the accretion of
gas-phase metals, and dust destruction in SN shocks are yet to be included in the
porosity evolution model.
For the evolution of grain size distribution, HA19 provide a comprehensive framework
that includes not only coagulation and shattering but also other relevant processes
for dust evolution.
Since the grain porosity is strongly related to the evolution of grain size distribution,
extending HA19's formulation to include the porosity evolution is a viable way
to predict the evolution of grain porosity in the entire history of a galaxy.

\citet{Voshchinnikov:2005aa,Voshchinnikov:2006aa} investigated how the extinction curve is affected
by porosity. This provides a useful hint for constraining
the grain porosity through comparison with observations.
Thus, we also predict extinction curves based on the grain size distribution and porosity
calculated by our newly developed model in this paper.
As a consequence, we will be able to address how the inclusion of porosity affects the
predicted extinction curves, and to provide a way to test the porosity evolution against the
observed extinction curves such as the MW extinction curve.
%%Further comparison with the MW extinction curve is also an important step for
%%examining if the effect of porosity calculated in a model is observationally acceptable or not.
Previously \citet{Voshchinnikov:2006aa} included porous silicates together with
compact graphite to fit the MW extinction curve. They succeeded in finding a fitting solution to
the MW extinction curve. However, in previous models (including 
\citealt{Mathis:1996aa} mentioned above), the porosity is a free parameter independent of
the grain radius.
%%and the grain size distributions have some free parameters not related to the porosity.
Thus, it is still necessary to test the grain-size-dependent porosity
realized as a result of the dust evolution (especially the evolution of grain size distribution)
against the MW extinction curve.
Moreover, we aim at treating the porosity and the grain size distribution {simultaneously}, so that
{both of them} are predicted quantities, not freely adjusted ones.

We note that dust emission SEDs \citep{Dwek:1997aa}
and starlight polarization \citep{Draine:2021aa} may also be useful to
constrain the porosity. Since predictions of these quantities require different
extensive frameworks, we leave them for future studies, and concentrate on
extinction curves for the first observational test.
Since the MW extinction curve is
best constrained at UV--near-IR wavelengths \citep[e.g.][]{Fitzpatrick:2007aa,Nozawa:2013aa},
we focus on
this wavelength range.
For the convenience of discussion, we divide the UV wavelength range into \textit{far-UV} and
\textit{near-UV} separated at $1/\lambda =6~\micron^{-1}$.
%%The IR properties of dust grains are tested in a future work together with the dust emission SEDs.

The goal of this paper is to clarify the evolution of porous grains
%%in a consistent manner with the dust evolution
in the entire history of a galaxy
{by extending H21's framework (i.e.\ treating coagulation as a source of porosity)
to include other processes and a longer evolutionary
time comparable to the present galaxy ages.} This enables us
to predict the grain porosity and its dependence on the grain radius
at various epochs in galaxy evolution. In other words, this study predicts for the first time the
grain-radius-dependent porosity achieved as a result of galaxy evolution.
We also calculate extinction curves to make clear how the evolution of porosity affects
the observed dust properties. A particular emphasis is put on the comparison with
the MW extinction curve.

This paper is organized as follows.
In Section~\ref{sec:model}, we formulate the evolution of grain size distribution and porosity
by developing HA19 and H21's frameworks.
We show the results in Section~\ref{sec:result}.
We make an additional effort of applying our results to the MW extinction curve in
Section \ref{sec:MW}. Based on the results, we provide further discussions
in Section \ref{sec:discussion}, and finally
give our conclusions in Section \ref{sec:conclusion}.

\section{Model}\label{sec:model}

We construct a model that describes the evolution of grain size distribution
and porosity in a galaxy.
We follow H21 for the definitions of the terms related to porous grains.
The \textit{filling factor} of a grain is defined as the
volume fraction occupied by the grain-composing material. The rest,
namely the volume fraction of vacuum,
is referred to as the \textit{porosity}. A grain is \textit{compact} if the filling factor is unity
(i.e.\ no porosity).

We consider all the processes included by HA19:
dust condensation in stellar ejecta (stellar dust production),
dust growth by the accretion of gas-phase metals,
dust destruction in SN shocks, grain growth by coagulation, and
grain disruption by shattering. Accretion and coagulation occur in the dense ISM, while
shattering takes place in the diffuse ISM.
We only consider the processes treated in HA19, and we neglect
other potentially important processes such as rotational disruption \citep{Hoang:2019aa},
which also depends on quantities (e.g.\ radiation field intensities)
not included in our model. 
To avoid the complexity arising from compound species,
we consider a single dust species in calculating the evolution of grain
size distribution and porosity.

{We calculate the dust evolution in a galaxy with a one-zone model.}
%%treat the galaxy as a one-zone object.
Thus, we neglect the spatial variety
of grain size distribution within the galaxy (or the grain size distribution can be
regarded as being averaged in the galaxy). Nevertheless, we still need to treat the
difference between the dense and diffuse ISM since some processes occur in one of
these phases as mentioned above. Thus, we introduce the dense gas fraction,
$\eta_\mathrm{dense}$, which decides the weight for the processes that only occur
in the dense ISM (see Section~\ref{subsec:param} for details).
{For each process, we assume a homogeneous medium and ignore
effects caused by small-scale inhomogeneity such as
ununiform gas density and dust abundance induced by supersonic
turbulence \citep{Hopkins:2016aa,Mattsson:2020aa,Li:2020ab,Li:2021aa}.}

For simplicity, we neglect the electric charge of grains. As long as the grain
motion is driven by turbulence as assumed in this paper, it is not likely that
the shattering rate is significantly affected by grain charge because of large grain
velocities (H21).
For coagulation and accretion, the charge of small grains could be important
but the grain charging in the dense ISM is very sensitive
to the assumed physical conditions \citep[e.g.][]{Ivlev:2015aa}.
In coagulation, large ($\gtrsim 0.01~\micron$)
grains still have large velocities so that the Coulomb barrier could be
neglected in collisions with large grains.
Negative charging for small grains could even
enhance dust growth by accretion \citep{Zhukovska:2016aa}, which
changes the grain radius and possibly the charging state as well.
Because
of the uncertainty in the grain charging,  we simply neglect
grain charge in this paper (as also done in HA19 and H21).

\subsection{Basic framework}\label{subsec:setup}

We formulate the evolution of grain size distribution and grain porosity
using the distribution functions of grain mass ($m$) and grain volume ($V$) at time $t$.
To treat porous grains, we introduce the
following two types of grain radius: characteristic radius ($a_\mathrm{ch}$) and mass-equivalent
radius ($a_m$). The characteristic radius, as defined by \citet{Okuzumi:2009aa},
is related to the grain volume (including vacuum) as
$V=(4\upi /3)a_\mathrm{ch}^3$, and the mass-equivalent radius is linked to the
grain mass as $m=(4\upi /3)a_m^3s$, where $s$ represents the bulk material density.
The filling factor, denoted as $\phi_m$, is expressed as
$\phi_m\equiv (a_m/a_\mathrm{ch})^3 (\leq 1)$ (note that $a_\mathrm{ch}>a_m$ for
porous grains), while the porosity is $1-\phi_m$ \citep[see also][]{Ormel:2009aa}.
For compact grains, $a_\mathrm{ch}=a_m$ or equivalently $m=sV$.
Since the variation of $s$ in a reasonable range does not affect our conclusion significantly,
we adopt the value of {astronomical
silicate (hereafter silicate;} $s=3.5$ g~cm$^{-3}$; \citealt{Weingartner:2001aa}) for the
calculations of grain size distribution and grain porosity.
%%We vary some parameters relevant for porosity, which are more important in this paper
%%(Section \ref{subsec:param}). Thus, we simply use the same parameters as in HM20 for
%%the material properties (specifically $s$ and $Q_\mathrm{D}^\star$ introduced later)
%%that do not directly affect the porosity evolution.

Here we introduce some quantities used to construct the basic equations.
Because there is ambiguity in the grain size (i.e.\ $a_m$ or $a_\mathrm{ch}$),
we use the grain mass distribution instead of the grain size distribution.
We define the distribution function of $m$ and $V$ at time $t$ as
the number density of dust grains in the $\mathbb{R}^3\times (m,\, V)$ space
($\mathbb{R}^3$ is a real 3-dimensional volume in the galaxy, for which we assume uniformity
because of the one-zone treatment), and denote it as $f(m,\, V,\, t)$.
We do not directly solve for $f(m,\, V,\, t)$ but use moment equations to save the computational cost.
The grain mass distribution at time $t$, $\tilde{n}(m,\, t)$ is introduced by the
zeroth moment of the above distribution function for $V$:
\begin{align}
\tilde{n}(m,\, t)\equiv\int_0^\infty f(m,\, V,\, t)\,\mathrm{d}V.\label{eq:n}
\end{align}
We note that $\tilde{n}(m,\, t)\,\mathrm{d}m$ is the number density of grains whose mass is between
$m$ and $m+\mathrm{d}m$.
We also use the first moment of $f$ for $V$:
\begin{align}
\bar{V}(m,\, t)\equiv\frac{1}{\tilde{n}(m,\, t)}\int_0^\infty Vf(m,\, V,\, t)\,\mathrm{d}V,\label{eq:V}
\end{align}
which is the mean volume of grains with mass $m$.
We also define
\begin{align}
\varrho (m,\, t) &\equiv m\tilde{n}(m,\, t),\label{eq:rho_def}\\
\psi (m,\, t) &\equiv \bar{V}(m,\, t)\tilde{n}(m,\, t).\label{eq:psi_def}
\end{align}
The set of $\varrho$ and $\psi$ contains the information on the distribution function of
grain mass and volume.

We construct the basic equations for $\varrho$ and $\psi$, which are equivalent to
the moments, $\tilde{n}$ and $\bar{V}$. In other words, we solve the moment equations.
To save the computational cost,
we further adopt the volume-averaging approximation \citep{Okuzumi:2009aa} (see also
Appendix~\ref{app:derivation} and H21), in which we represent the volume by
$\bar{V}$; that is, all grains with the same mass are approximated to have the same volume.
Note that
\begin{align}
\bar{V}(m,\, t)=m\psi (m,\, t)/\varrho (m,\, t)\label{eq:V_close}
\end{align}
holds from equations (\ref{eq:rho_def}) and (\ref{eq:psi_def}), so that the two differential equations for
$\varrho$ and $\psi$ are closed.
Because of the volume-averaging approximation, the filling factor reduces to a function of $m$
(at each $t$) as
\begin{align}
\phi_m=\frac{m}{s\bar{V}}.\label{eq:filling}
\end{align}

The dust mass density (per volume) is the integration of $\varrho(m,\, t)$ for $m$, so that
the dust-to-gas mass ratio at time $t$, $\mathcal{D}(t)$, is written as
\begin{align}
\mu_\mathrm{H}m_\mathrm{H}n_\mathrm{H}\mathcal{D}(t)=\int_0^\infty\varrho (m,\, t)\,\mathrm{d}m,
\label{eq:dg}
\end{align}
where $\mu_\mathrm{H}=1.4$ is the gas mass per hydrogen, $m_\mathrm{H}$ is
the hydrogen atom mass, and $n_\mathrm{H}$ is the hydrogen
number density. Note that the gas density, $\rho_\mathrm{gas}$, is estimated as
$\rho_\mathrm{gas}=\mu_\mathrm{H}m_\mathrm{H}n_\mathrm{H}$.

In the following subsections, we present the basic equations for $\varrho (m,\, t)$ and
$\psi (m,\, t)$. For convenience, we formulate individual processes separately;
in reality, we solve all processes together at each time-step.
The continuous differential equations given below are solved by
discretizing the entire grain radius range
($a_m=3\times 10^{-4}$--10 $\micron$) into 128
grid points with logarithmically equal spacing.
The discretization algorithm is described in appendix B of HA19.
We set ${\varrho}_\mathrm{d}(m,\, t)=0$ and $\psi_\mathrm{d}(m,\, t)=0$
at the maximum and minimum grain radii for
the boundary conditions.
The integration in the range $[0,\,\infty ]$ is practically performed between
the minimum and maximum grain masses ($m_\mathrm{min}$ and $m_\mathrm{max}$,
respectively) corresponding to the above minimum and
maximum $a_m$, respectively (or equivalently we regard $\varrho$ and $\psi$ out of the above
grain mass range as zero).

\subsection{Stellar dust production}

A certain fraction of the metals ejected from SNe and AGB stars
are condensed into dust
\citep[e.g.][]{Kozasa:1989aa,Todini:2001aa,Nozawa:2003aa,Ferrarotti:2006aa,Ventura:2014aa}.
Although the dust condensation efficiency in the stellar ejecta could vary depending on the
progenitor star mass, the dependence is uncertain
\citep[][and references therein]{Inoue:2011aa,Kuo:2013aa}.
Therefore, following HA19, we adopt a
constant parameter ($f_\mathrm{in}$) that describes
the condensation efficiency of metals in stellar ejecta.

We write the change of the grain
size distribution by stellar dust production as
\begin{align}
\left[\frac{\upartial\varrho (m,\, t)}{\upartial t}\right]_\mathrm{star}=
f_\mathrm{in}\varrho_\mathrm{gas}\dot{Z}\, m\tilde{\varphi} (m),\label{eq:stellar}
\end{align}
where $Z$ is the metallicity (including both gas and dust phases)
with the dot meaning the increasing rate, and
$m\tilde{\varphi} (m)$ is the mass distribution function of
the dust grains produced by stars. The normalization of $m\tilde{\varphi} (m)$
is determined so that the integration
for the whole grain mass range is unity.
We adopt $f_\mathrm{in}=0.1$ (HA19).
It is often convenient to define the grain
size distribution corresponding to $\tilde{\varphi} (m)$ as
$\varphi (a)\,\mathrm{d}a\equiv\tilde{\varphi}(m)\,\mathrm{d}m$.
We assume the following lognormal
form for ${\varphi}(a)$:
\begin{align}
{\varphi}(a)=\frac{C_\varphi}{a}\exp\left\{
-\frac{[\ln (a/a_0)]^2}{2\sigma^2}\right\} ,
\end{align}
where $C_\varphi$ is the normalization factor, $\sigma$ is the standard deviation,
and $a_0$ is the central
grain radius ($\sigma =0.47$ and $a_0=0.1~\micron$ following \citealt{Asano:2013aa};
HA19). This functional
form assumes that stars produce large ($\sim 0.1~\micron$) grains
(see the arguments in e.g.\ \citealt{Nozawa:2007aa} and
\citealt{Yasuda:2012aa} for the typical grain radii condensed in SNe and in AGB star winds,
respectively).
For the chemical evolution (evolution of $Z$), we adopt a simple functional form later
in Section \ref{subsec:param}.

We assume that dust condensation in stellar ejecta occurs predominantly
through the accretion of atoms or monomers (i.e.\
growth at an atomic level, not through the attachment of macroscopic grains).
In atomic-level condensation, we do not expect that porosity develops.
However, as shown by \citet{Sarangi:2015aa}, coagulation can take place in SN ejecta
\citep[see also][]{Sluder:2018aa}
although its efficiency depends on the clumpiness of the ejecta.
%%if grain--grain collisions occur frequently in SNe and/or AGB star winds,
%%coagulation may create porous grains. In particular, if the gas is turbulent in dust-condensation
%%regions, relative velocities among grains could emerge, leading to grain--grain collisions.
Coagulation is also included in a dust condensation calculation for AGB stars
\citep{Gobrecht:2016aa}; however, whether coagulation could play a significant
role compared with dust condensation (nucleation and accretion) is not clear.
Since the general importance of coagulation in stellar ejecta is still uncertain, we simply
assume that stardust grains are compact.
This treatment gives a conservative estimate of the porosity evolution, and also serves to focus on the
creation of porosity through interstellar processing.
We note that pre-solar grains originating from SNe and AGB stars found in meteorites are
compact \citep[e.g.][]{Anders:1993aa,Amari:1994aa}.

The compactness of the grains formed in stellar ejecta leads to the following expression for
the change of the volume-weighted distribution function by stellar dust production:
\begin{align}
\left[\frac{\upartial\psi (m,\, t)}{\upartial t}\right]_\mathrm{star}=\frac{1}{s}
\left[\frac{\upartial\varrho (m,\, t)}{\upartial t}\right]_\mathrm{star},
\end{align}
which is based on the relation $\phi_m=\varrho/(s\psi )$ with $\phi_m=1$
(equations \ref{eq:V_close} and \ref{eq:filling}).

\subsection{Coagulation and shattering}\label{subsec:coag}

The evolution of grain size distribution and porosity in shattering and coagulation
was already formulated by H21. For compact grains, the equations for the evolution
of grain size distribution had already been
developed \citep{Jones:1994aa,Jones:1996aa,Hirashita:2009ab}. We use
the moment equations of 2-dimensional ($m,\, V$) Smolchowski equation for coagulation
with the volume-averaging approximation based on
\citet{Okuzumi:2009aa,Okuzumi:2012aa},
{who investigated how coagulation develops grain porosity.}
H21 extended their equation to describe shattering.
Below we only describe the final equations we use in this paper, and refer the interested
reader to H21 for the derivation.

The time evolution of the two distribution functions, $\varrho$ and $\psi$, is described
by the following equations with the subscripts `coag' and `shat' indicating
coagulation and shattering, respectively:
\begin{align}
\lefteqn{\left[\frac{\upartial\varrho (m,\, t)}{\upartial t}\right]_\mathrm{coag/shat}
= -m\varrho (m,\, t)\int_0^\infty
\frac{K_{m,m_1}}{mm_1}\varrho (m_1,\, t)\mathrm{d}m_1}\nonumber\\
& +\int_0^\infty\int_0^\infty\frac{K_{m_1,m_2}}{m_1m_2}\varrho (m_1,\, t)\varrho(m_2,\, t)
m\bar{\theta} (m;\, m_1,\, m_2)\mathrm{d}m_1\mathrm{d}m_2,\label{eq:rho}
\end{align}
\begin{align}
\lefteqn{\left[\frac{\upartial\psi (m,\, t)}{\upartial t}\right]_\mathrm{coag/shat}}\nonumber\\
&=  -\bar{V}(m,\, t)\psi (m,\, t)\int_0^\infty
\frac{K_{m,m_1}}{\bar{V}(m,\, t)\bar{V}(m_1,\, t)}\psi (m_1,\, t)\mathrm{d}m_1\nonumber\\
& +\int_0^\infty\int_0^\infty\frac{K_{m_1,m_2}}{\bar{V}(m_1)\bar{V}(m_2)}
\psi (m_1,\, t)\psi(m_2,\, t)
(V_{1+2})_{m_1,m_2}^m\nonumber\\
&\times\bar{\theta} (m;\, m_1,\, m_2)\mathrm{d}m_1\mathrm{d}m_2,
\label{eq:psi}
\end{align}
where $K_{m_1,m_2}$ is the Kernel function in the collision between grains with mass $m_1$ and
$m_2$ (referred to as grains 1 and 2, respectively),
$\tilde{\theta}(m;\, m_1,\, m_2)$ is the distribution function of
grain mass produced from grain 1 in the collision between grains 1 and 2, and
$(V_{1+2})_{m_1,m_2}^m$ is the volume of the newly produced grain with mass $m$.
%%We note that a factor 1/2 in front of the second term on the right-hand size in the above two
%%equations is not necessary because we count the collisional products originating from
%%$m_1$ and $m_2$ separately. This is why we do not have a factor 1/2 in front
%%of the second term on the right-hand side in equations (\ref{eq:rho}) and (\ref{eq:psi})
%%(cf.\ \citealt{Okuzumi:2009aa}).

The collision kernel for the collision between grains 1 and 2 is estimated as
\begin{align}
K_{m_1,m_2}=\sigma_{1,2}v_{1,2},
\end{align}
where $\sigma_{1,2}=\upi (a_\mathrm{ch1}+a_\mathrm{ch2})^2$
($a_\mathrm{ch1}$ and $a_\mathrm{ch2}$ are characteristic radii of grain 1 and 2,
respectively) is the collisional cross-section,
and $v_{1,2}$ is the relative velocity between grains 1 and 2.
We assume that the grain velocity is induced by {interstelalr}
turbulence, and the resulting typical
velocity as a function of grain radius is described as
(\citealt{Ormel:2009aa}; H21)
\begin{align}
v_\mathrm{gr}(m) &= 1.1\mathcal{M}^{3/2}\phi_m^{1/3}\left(
\frac{a_m}{0.1~\micron}\right)^{1/2}\left(\frac{T_\mathrm{gas}}{10^4~\mathrm{K}}\right)^{1/4}
\left(\frac{n_\mathrm{H}}{1~\mathrm{cm}^{-3}}
\right)^{-1/4}\nonumber\\
&\times \left(\frac{s}{3.5~\mathrm{g~cm}^{-3}}\right)^{1/2}~\mathrm{km~s}^{-1},
\label{eq:vel}
\end{align}
where $\mathcal{M}$ is the Mach number of the largest-eddy velocity (practically used to
normalize the grain velocity), and
$T_\mathrm{gas}$ is the gas temperature.
We apply $\mathcal{M}=3$ for shattering and $\mathcal{M}=1$ for coagulation
\citep{Hirashita:2020aa} to obtain similar grain velocities to those in \citet{Yan:2004aa}.
We also fix $s=3.5$ g cm$^{-3}$ as mentioned above. The gas density and temperature are set in
Section \ref{subsec:param}.
We estimate the relative velocity $v_{1,2}$ between grains 1 and 2 by assuming a random direction in every
calculation of the collision kernel \citep{Hirashita:2013aa}.
{We neglect grain--grain collisions in other environments such as supernova shocks
\citep{Jones:1996aa,Kirchschlager:2019aa} for simplicity and note that additional shattering
and coagulation
cannot be distinguished from these processes associated with interstellar turbulence in our model.}

We refer the reader to H21 (see their section 2.3) for the treatment of collisional products.
We only give summaries in what follows.

\subsubsection{Collisional products in coagulation}\label{subsubsec:coag}

The porosity of a coagulated grain in a grain--grain collision
is broadly determined by the ratio between the impact energy and the rolling energy
\citep{Okuzumi:2012aa}. H21 kept this physical essence, but somewhat simplified the
formulae that describe the porosity of collisional products.
We here summarize H21's treatment of porosity, focusing on the physical essence
without repeating the full equations.

The rolling energy, denoted as $E_\mathrm{roll}$, is the energy necessary
for a monomer to roll over 90 degrees on the surface
of another monomer \citep{Dominik:1997aa,Wada:2007aa}.
The rolling energy is written as $E_\mathrm{roll}=12\upi^2\gamma R_{1,2}\xi_\mathrm{crit}$,
where $\gamma$ is the surface energy per unit contact area, $R_{1,2}$ is the reduced grain radius
[assumed to be equal to the reduced radius of the two grains:
$a_{m_1}a_{m_2}/(a_{m_1}+a_{m_2})$; H21], and $\xi_\mathrm{crit}$ is
the critical displacement of rolling (beyond which particle motions enter
the inelastic regime).
We estimate the volume of the coagulated grain ($V_{1+2}$) from grains 1 and 2
(with volumes $V_1$ and $V_2$, respectively). The volume of the coagulated grain depends
on the impact (kinetic) energy, $E_\mathrm{imp}$, relative to $E_\mathrm{roll}$
(see H21 for the actual equations we adopted):
(i) $V_{1+2}\simeq V_1+V_2+V_\mathrm{void}$
for $E_\mathrm{imp}\ll E_\mathrm{roll}$ (reflecting the hit-and-stick regime).
(ii) The newly created volume ($V_\mathrm{void}$) is compressed
if $E_\mathrm{imp}\gtrsim E_\mathrm{roll}$. (iii) If $E_\mathrm{imp}$ becomes comparable
to or higher than $n_\mathrm{c}E_\mathrm{roll}$, the grain is compressed further,
approaching $V_{1+2}\simeq (1+\epsilon_V)(V_\mathrm{1,comp}+V_\mathrm{2,comp})$
at $E_\mathrm{imp}\gg n_\mathrm{c}E_\mathrm{roll}$, where
$V_{i,\mathrm{comp}}\equiv m_i/s$ (for $i=1$ and 2), $n_\mathrm{c}$ is the number of contact
points (treated as a fixed parameter), and $\epsilon_V$ is the parameter that regulates the
maximum compression.
The coagulated grain has a mass of $m_1+m_2$, and it is put in the appropriate
grain mass bin.

Since H21 already investigated detailed parameter dependence for coagulation, we basically
fix the parameter values necessary for the calculation of coagulation.
See H21 for the discussions on the meaning and choice of each parameter.
We adopt $\gamma =25$ erg cm$^{-2}$ (value for silicate, originally for quartz;
\citealt{Chokshi:1993aa}; note that, as mentioned above, we adopt the silicate properties for
the calculation of grain size distribution and porosity),
$\zeta_\mathrm{crit}=10$ \AA, $\epsilon_V=0.5$ and $n_\mathrm{c}=30$ unless
otherwise stated.
We should also keep in mind that
the surface energy of dry silica may be larger than the above values
\citep{Kimura:2015aa,Steinpilz:2019aa}. This implies that
$E_\mathrm{roll}$ is underestimated. Graphite and water ice may have larger $E_\mathrm{roll}$.
To examine a possibility of larger $E_\mathrm{roll}$,
we also examine a higher $E_\mathrm{roll}$ with $\gamma =100$ erg cm$^{-2}$
(appropriate for water ice; \citealt{Israelachvili:1992aa,Wada:2007aa}, noting that graphite
has an intermediate value (75 erg cm$^{-2}$; \citealt{Dominik:1997aa})
and $\zeta_\mathrm{crit}=30$ \AA\ (these are the largest values examined in H21 based on
\citealt{Dominik:1997aa,Heim:1999aa,Wada:2007aa})
in Section \ref{subsec:improve}.

\subsubsection{Collisional products in shattering}\label{subsubsec:shat}

The total ejected mass ($m_\mathrm{ej}$)
of shattered fragments from the original grain $m_1$ is determined by the ratio
between the impact energy per grain mass (specific impact energy)
and the specific energy necessary for the catastrophic disruption
($Q_\mathrm{D}^\star$) \citep{Kobayashi:2010aa}.
The catastrophic disruption is defined as half of the grain mass being shattered;
that is, $m_\mathrm{ej}=m_1/2$.
If the specific impact energy is much smaller (larger) than $Q_\mathrm{D}^\star$,
$m_\mathrm{ej}$ is negligible (comparable to $m_1$).
We adopt $Q_\mathrm{D}^\star = 4.3\times 10^{10}$ erg g$^{-1}$, which is valid for
silicate (HA19). If we adopt the value of graphite, which is 5 times smaller,
shattering produces larger numbers of fragments, leading to an increase of porosity
(Section~\ref{subsec:improve}).
The total ejected mass
$m_\mathrm{ej}$ is distributed into fragments, of which the size distribution is
assumed to be a power-law with an index of
$-3.3$ \citep{Jones:1996aa}. The remnant ($m_1-m_\mathrm{ej}$) is put in the appropriate
grain radius bin.
The maximum and minimum masses of the fragments
are assumed to be
$m_\mathrm{f,max}=0.02m_\mathrm{ej}$ and
$m_\mathrm{f,min}=10^{-6}m_\mathrm{f,max}$, respectively \citep{Guillet:2011aa};
if $m_\mathrm{f,min}$ is smaller than the minimum grain mass (corresponding
$a_m=3$ \AA), we apply $m_\mathrm{f,min}=m_\mathrm{min}$.
We remove grains if the mass-equivalent grain radius becomes smaller
than 3 \AA\ (this happens when the maximum fragment mass is less than $m_\mathrm{min}$).

The remnant can be compressed after the collision.
H21 considered two cases: one is that we neglect this compression, and the other is
that we assume compaction of a volume equivalent to $m_\mathrm{ej}$. We adopt
the latter case in this paper,
since it gives a more conservative estimates for the porosity
(however, this compaction only affects the grains at $a_m\gtrsim 0.1~\micron$,
and its effect is minor compared with the change of $\gamma$ and
$\xi_\mathrm{crit}$).
Thus, we assume that the volume fraction equal to $m_\mathrm{ej}/m_1$ becomes
compact after the collision. We also set a limit of $\phi_m\leq 1$ to avoid the compression
proceeding beyond that point.

\subsection{Dust destruction by SN shocks and dust growth by accretion}\label{subsec:dest_acc}

Here we consider the evolution of grain size distribution and porosity through
dust destruction by SN shocks and dust growth by the accretion of gas-phase metals.
The evolution of grain size distribution by these two processes can be described
by a
conservation law of the number of grains (HA19). As a consequence, the evolution is described
by an advection equation in the grain radius (or grain mass) space.
In this paper, we extend the equations to include the grain volume.
As derived in Appendix \ref{app:derivation}, the evolution of $\varrho$ and $\phi$
through the above two processes is described by
\begin{align}
\left[\frac{\upartial\varrho (m,\, t)}{\upartial t}\right]_\mathrm{dest/acc}
&= -\frac{\upartial}{\upartial m}[\dot{m}\varrho (m,\, t)]+
\frac{\dot{m}}{m}\varrho (m,\, t),\label{eq:dest_acc_m}\\
\left[\frac{\upartial\psi (m,\, t)}{\upartial t}\right]_\mathrm{dest/acc}
&= -\frac{\upartial}{\upartial m}[\dot{m}\psi (m,\, t)]+
\frac{\dot{V}}{\bar{V}}\psi (m,\, t),\label{eq:dest_acc_V}
\end{align}
where $\dot{m}\equiv \mathrm{d}m/\mathrm{d}t$, $\dot{V}\equiv \mathrm{d}V/\mathrm{d}t$,
and $\bar{V}$ are functions of $m$ and $t$
(Appendix \ref{app:derivation}). The subscripts `dest' and `acc' indicate
dust destruction and accretion, respectively. The above equations indicate that
$\varrho$ and $\psi$ follow the same `advection' equations except for the factor
$\dot{m}/m$ or $\dot{V}/\bar{V}$ in the source term. This means that the difference
between the growth rates of mass and volume changes the porosity. In what follows,
we evaluate $\dot{m}$ and $\dot{V}$ for each process.

\subsubsection{Dust destruction by SN shocks}\label{subsubsec:destruction}

Since our model is not capable of treating the inhomogeneity within a grain,
we simply assume that the mass and volume are destroyed at the same rate
in SN destruction. This is correct if the material and vacuum are mixed homogeneously
within a grain. Based on this assumption, we write $\dot{m}$ and $\dot{V}$ as
\begin{align}
\dot{m} &= -m/\tau_\mathrm{dest}(m,\,\bar{V}),\\
\dot{V} &= -\bar{V}/\tau_\mathrm{dest}(m,\,\bar{V}),
\end{align}
where $\tau_\mathrm{dest}(m,\,\bar{V})$ is the destruction time-scale as
a function of grain mass and volume, which is given below.
{Since $\dot{m}/m=\dot{V}/\bar{V}$, the filling factor (porosity) does not change
in dust destruction in our model.}

For the destruction time-scale, we basically adopt the functional form from HA19
\citep[originally from][]{McKee:1989aa}
with a modification for porosity.
The time-scale on which the ISM is swept once on average by SNe
is estimated by $\tau_\mathrm{sw}\equiv M_\mathrm{gas}/(M_\mathrm{s}\gamma )$,
where $M_\mathrm{gas}$ is the total gas mass of the ISM,
$M_\mathrm{s}$ is the gas mass swept by a single SN blast,
and $\gamma$ is the SN rate.
Using the sweeping time-scale, we evaluate the destruction time-scale as
\begin{align}
\tau_\mathrm{dest}(m)=
\frac{\tau_\mathrm{sw}\phi_m^{2/3}}{\epsilon_\mathrm{dest}(m)},
\label{eq:tau_dest}
\end{align}
where 
$\epsilon_\mathrm{dest}(m)$ is the fraction of destroyed dust in a single
passage of SN shock as a function of grain mass, and $\phi_m^{2/3}$ represents
the effective increase of the grain surface area (explained below). We adopt
the following functional form for the efficiency (HA19):
$\epsilon_\mathrm{dest}(m)=1-\exp [-0.1({a_m}/{0.1~\micron})^{-1}]$.
The SN rate $\gamma$ is tightly coupled with the chemical enrichment and is given later
in Section \ref{subsec:param}
together with $M_\mathrm{s}$.

The destruction occurs by sputtering, which is a surface process. Thus, the
destruction rate is proportional to the the surface-to-volume ratio. To be precise,
the surface-to-volume
ratio of fluffy grain depends on the fractal dimension \citep[e.g.][]{Okuzumi:2009aa}; in this paper, however,
we simply assume that the porosity increases the surface by a factor of
$(a_\mathrm{c}/a_m)^2=\phi_m^{-2/3}$, which is regarded as the effective increase of the projected
area of a grain. Thus, the dust destruction time-scale is
assumed to be proportional to $\phi_m^{2/3}$.

{As mentioned in Section \ref{subsec:coag}, we neglect shattering associated with
SN shocks, although shattering enhances the dust destruction by sputtering
owing to the resulting smaller grain sizes \citep{Kirchschlager:2021aa}. We also ignore
the metallicity dependence of
SN destruction efficiency, which could have a significant imprint on the
evolution of dust abundance \citep{Yamasawa:2011aa,Priestley:2021aa}. The metallicity
dependence of SN destruction could delay/enhance
the increase of grain abundance at low/high metallicity.
However, SN dust destruction has a minor influence on
the porosity compared with coagulation and shattering,
so that the detailed treatment of
SN destruction does not influence the conclusion (Section \ref{subsec:uncertainties}).}

\subsubsection{Dust growth by accretion}\label{subsubsec:accretion}

For accretion, the growth rate $\dot{m}$ is
estimated as
\begin{align}
\dot{m} &= \xi (t)m/\tau_\mathrm{acc} (m,\,\bar{V}),\label{eq:acc_m}\\
\dot{V}  &= \dot{m}/s,\label{eq:acc_V}
\end{align}
where $\xi (t)\equiv 1-\mathcal{D}(t)/Z (t)$
is the fraction of metals in the gas phase
and $\tau_\mathrm{acc} (m,\,\bar{V})$ is the accretion (dust growth) time-scale given later.
We assume that the newly increased volume by accretion is
compact (equation \ref{eq:acc_V}); that is, new condensation of material
creates compact solid (like dust condensation in stellar ejecta).
We adopt the following accretion time-scale
(HA19):\footnote{The sticking efficiency $S$ is fixed to 0.3.}
\begin{align}
\tau_\mathrm{acc}=  \tau_0\phi_m^{2/3}\left(\frac{a_m}{0.1~\micron}
\right)\left(\frac{Z}{\mathrm{Z}_{\sun}}\right)^{-1}
\left(\frac{n_\mathrm{H}}{10^3~\mathrm{cm}^{-3}}
\right)^{-1}\left(\frac{T_\mathrm{gas}}{10~\mathrm{K}}
\right)^{-1/2},
\label{eq:tau_acc}
\end{align}
where $\tau_0$ is a constant given below,
$Z_{\sun}$ is the solar metallicity (we adopt the same solar metallicity
$Z_{\sun}=0.02$ as in HA19). Because we fix $n_\mathrm{H}=10^3$ cm$^{-3}$
and $T_\mathrm{gas}=10$ K (Section \ref{subsec:param})
in the dense ISM where accretion occurs, $\tau_\mathrm{acc}$
is a function of $a$ and $Z$. We adopt $\tau_\mathrm{0}=5.37\times 10^7$ yr (HA19).
Since accretion is a surface process, the factor $\phi_m^{2/3}$ is multiplied
to take into account the effect of porosity on the grain surface area
as also done in equation (\ref{eq:tau_dest}).

For accretion, $\dot{V}/\bar{V}=\dot{m}/(s\bar{V})=(\dot{m}/m)[m/(s\bar{V})]=\phi_m(\dot{m}/m)\leq\dot{m}/m$
using equations (\ref{eq:filling}), (\ref{eq:acc_m}), and (\ref{eq:acc_V}).
This means that the volume grows less rapidly than the mass. Thus, the porosity decreases
(or the filling factor increases) as a result of accretion. This is {due to our assumption that
newly condensed portion of a grain is compact.}

\subsection{All processes together}\label{subsec:param}

As mentioned in the beginning of this section, some processes occur only in a limited
ISM phase. Coagulation and accretion take place in the dense ISM, while shattering
occurs in the diffuse ISM.
For simplicity, we divide the ISM into the two phases: the dense and diffuse ISM.
We adopt
$n_\mathrm{H}=10^3$~cm$^{-3}$ and $T_\mathrm{gas}=10$~K (typical of molecular clouds) for the
dense ISM and
$n_\mathrm{H}=0.3$ cm$^{-3}$ and $T_\mathrm{gas}=10^4$ K for the diffuse ISM, following H21.

{Because of the one-zone nature of the model},
we adopt the common grain size distribution for the two ISM phases.
The contributions from each of the two ISM phases to the time variation of the grain size distribution
are added with an appropriate weight,
for which we use the dense gas fraction, $\eta_\mathrm{dense}$.
Although spatially resolved simulations showed that the grain size distributions are
different between the dense and diffuse ISM \citep{Aoyama:2020aa}, our approach of a single (averaged) grain size distribution
is expected to give a reasonable approximation (i) for a volume large enough
to contain a statistically significant mass of both ISM  phases and/or (ii) for a time much longer than
the mixing time-scale of the ISM phases ($\sim 10^7$~yr; e.g.\ \citealt{McKee:1989aa}).
Given that detailed modelling of
the ISM is beyond the scope of this paper, we simply fix $\eta_\mathrm{dense}$ as done
by \citet{Hirashita:2020aa} (or we regard
the constant value of $\eta_\mathrm{dense}$ as the average over the entire history of the
galaxy).
Since the two ISM phases have different gas densities,
it is convenient to define the distribution functions per hydrogen,
$\tilde{\varrho}\equiv\varrho /n_\mathrm{H}$ and
$\tilde{\psi}\equiv\psi /n_\mathrm{H}$, which are independent of the density scaling.
We calculate the evolution of $\tilde{\varrho}$ and $\tilde{\psi}$ by
\begin{align}
\frac{\upartial\tilde{\varrho}(m,\, t)}{\partial t} &=\sum_i\frac{1}{n_{\mathrm{H},i}}
\left[\frac{\upartial{\varrho}(m,\, t)}{\partial t}\right]_i\eta_i,
\label{eq:rho_mix}\\
\frac{\upartial\tilde{\psi}(m,\, t)}{\partial t} &=\sum_i\frac{1}{n_{\mathrm{H},i}}
\left[\frac{\upartial{\psi}(m,\, t)}{\partial t}\right]_i\eta_i,
\label{eq:psi_mix}
\end{align}
where the subscript $i$ indicates the process (`star', `coag', `shat', `dest', or `acc'),
$\eta_i$ is the weight for the ISM phase ($\eta_i=\eta_\mathrm{dense}$ for coagulation
and accretion, $\eta_i=1-\eta_\mathrm{dense}$ for shattering, and $\eta_i=1$ for
stellar dust production and dust destruction).
$n_{\mathrm{H},i}$ is the hydrogen number density in the appropriate ISM phase
($10^3$ cm$^{-3}$ for coagulation and accretion, 0.3 cm$^{-3}$ for shattering, and
arbitrary for stellar dust production and dust destruction; note that the equations
for the last two processes have a linear dependence on the density).
As is clear from the above equations, $\eta_\mathrm{dense}$ is
practically a parameter that determines the balance between
dust growth (coagulation and accretion) and dust fragmentation
in our model.

We also need the evolution of metallicity $Z$ and SN rate $\gamma$.
Because the construction of a detailed chemical evolution model is not the purpose of this paper,
we adopt a simple functional form used by HA19:
$Z=0.6(t/\tau_\mathrm{SF})$ Z$_{\sun}$, where $\tau_\mathrm{SF}$
is the star formation time-scale (i.e.\ the star formation rate $\psi$ is given by
$\psi =M_\mathrm{gas}/\tau_\mathrm{SF}$).
Since there are no metals at $t=0$, we adopt $\varrho (m,\, t=0)=0$ and $\psi (m,\, t=0)=0$
for the initial condition.
The above metallicity evolution indeed fits the one in \citet[][see their fig.\ 6]{Asano:2013aa}.
For the SN rate, we only consider core-collapse SNe with progenitor mass $>8~\mathrm{M}_{\sun}$.
The lifetimes of such massive stars are short enough to be regarded as instantaneous,
so that $\gamma$ is proportional to $\psi$. Thus, the sweeping time-scale
$\tau_\mathrm{sw}$ defined in Section \ref{subsubsec:destruction} is evaluated as
$\tau_\mathrm{sw}=M_\mathrm{gas}/(M_\mathrm{s}\gamma )=
M_\mathrm{gas}/(\nu_\mathrm{SN}\psi M_\mathrm{s})=
\tau_\mathrm{SF}/(\nu_\mathrm{SN}M_\mathrm{s})$, where $\nu_\mathrm{SN}$ is the
proportionality constant between $\psi$ and $\gamma$. Adopting the \citet{Chabrier:2003aa}
initial mass function with a stellar mass range of 0.1--100~M$_{\sun}$, we estimate that
$\nu_\mathrm{SN}=1.0\times 10^{-2}$ M$_{\sun}^{-1}$.
We adopt $M_\mathrm{s}=6800$ M$_\odot$ \citep{McKee:1989aa,Nozawa:2006aa}, and
obtain $\tau_\mathrm{sw}=\tau_\mathrm{SF}/68$.

For the setup of galaxy parameters, we choose $\eta_\mathrm{dense}=0.5$ and
$\tau_\mathrm{SF}=5$ Gyr as fiducial values. We vary these two parameters to examine the
effects of the balance among various processes and of the metal enrichment time-scale.

\subsection{Calculation of extinction curves}\label{subsec:ext_method}

Following H21, we calculate extinction curves to predict an observable property.
We focus on two representative grain materials: silicate and carbonaceous species.
Note that we used the silicate properties for the calculations of grain size distribution and porosity.
Since the uncertainty in the parameters has as large an influence as the material
properties, we simply use the above grain size distribution and porosity for both materials.
Moreover, the comparison under a common grain size distribution is useful to observe the
difference purely caused by the grain material.

For the carbonaceous material,
as mentioned in H21, graphite has a problem that the central wavelength of the
2175~\AA\ bump shifts as porosity increases, which is not observed in the MW.
Other than the 2175 \AA\ bump,
the effects of porosity are similar between amorphous carbon (amC) and graphite. Thus, for the
carbonaceous species, we focus on amC.
\citet{Voshchinnikov:2006aa}, in their fitting to the MW extinction curve, treated graphite
as a compact component, while they treated silicate grains as porous.
Thus, we also discuss the 2175 \AA\ carrier as a separate component from our porosity
modelling.
We only summarize the calculation method of extinction curves and refer the interested reader to H21 for
further details.
For silicate and carbonaceous dust, we adopt the optical constants of astronomical silicate
from \citet{Weingartner:2001aa} and those of amC
from `ACAR' in \citet{Zubko:1996aa}, respectively.

The effective dielectric permittivity ($\bar{\varepsilon}$)
is calculated using the effective medium theory
%%(EMT)
with the Bruggeman mixing rule.
%% for a mixture of a grain material (either silicate or amC) and vacuum
%%using the Bruggeman mixing rule. The fraction of the vacuum is determined by the
%%porosity calculated above for each grain mass.
We separately calculate the extinction curves for silicate and amC
using the grain size distribution calculated above.
%%This mixing rule as well as the Garnett mixing rule gives reasonable extinctions as long as
%%the grains do not have substructures (e.g.\ monomers) larger than the wavelength
%%\citep{Voshchinnikov:2005aa}.
%%As shown later, the porous grains are formed by coagulation of grains
%%smaller than $\sim 0.1~\micron$ and we are interested in wavelengths longer than
%%0.1 $\micron$. Thus, the above condition is satisfied in this paper.
%%Other mixing rules such as the Garnett mixing rule
%%\citep[e.g.][]{Voshchinnikov:2005aa}.
%%Even if we model the inhomogeneity using
%%multi-layered spheres, the difference in the extinction is expected to be less than
%%$\sim 10$ per cent \citep{Voshchinnikov:2005aa,Shen:2008aa}.
%%Using the Bruggeman mixing rule, the averaged dielectric permittivity $\bar{\varepsilon}$
%%is obtained from
%%\begin{equation}
%%(1-\phi_m) \frac{1 - \bar{\varepsilon}}{1 + 2 \bar{\varepsilon}} + 
%%\phi_m \frac{\varepsilon_2 - \bar{\varepsilon}}{\varepsilon_2 + 2 \bar{\varepsilon}} = 0,
%%\end{equation}
%%where $\varepsilon_2$ is the dielectric permittivity of the material
%%(note that the first material is assumed to be vacuum so
%%$\varepsilon_1=1$).\footnote{{We adopt Gaussian-cgs units.}}
We assume each dust grain to be spherical with radius $a_\mathrm{ch}$ and refractive index
$\bar{m}=\sqrt{\bar{\varepsilon}}$.
The cross-section $C_{\mathrm{ext},m}$
is calculated by the Mie theory \citep[][]{Bohren:1983aa}.
To concentrate on the effect of porosity, we do not consider
more complicated structures,
such as compounds and grain mantles. Such structures may be important
if we aim at more detailed modelling of the MW extinction curve
\citep{Mathis:1996aa,Jones:2017aa}.

The extinction at wavelength $\lambda $, $A_{\lambda }$ (mag) is
calculated as
\begin{align} 
A_{\lambda}=(2.5\log_{10} \mathrm{e})L\displaystyle\int_{0}^{\infty}
\tilde{n}(m)\,C_{\mathrm{ext},m}\,\mathrm{d}m,
\end{align}
where $L$ is the path length.
We present the extinction per hydrogen $A_\lambda/N_\mathrm{H}$
($N_\mathrm{H}=n_\mathrm{H}L$ is the column density of
hydrogen nuclei), which scales with the dust abundance. We also
show
the normalized extinction $A_\lambda /A_V$ (the $V$ band wavelength corresponds to
$\lambda ^{-1}=1.8\,\mu {\rm m}^{-1}$) to focus on the wavelength dependence.
The path length $L$ is cancelled out in both expressions.

To clarify the effect of porosity, we calculate the \textit{extinction curve without porosity} by
forcing the filling factor of all grains to be unity (i.e.\ $\phi_m =1$).
The extinction calculated in this way is denoted as $A_{\lambda ,1}$.
The effect of porosity is quantified by
$A_\lambda /A_{\lambda ,1}$.

\section{Results}\label{sec:result}

We present the evolution of grain size distribution, filling factor and extinction curve
in this section. As shown by HA19, different processes are dominant
at different ages; thus, we expect that the effect of each process on the porosity will be
clear if we examine the time evolution. We also change $\eta_\mathrm{dense}$ and
$\tau_\mathrm{SF}$, which affect the balance among the processes.

For the presentations, the grain size distribution is shown in the form of
$a_m^4n(a_m)/n_\mathrm{H}$, where $n(a_m)$ is defined as
$n(a_m)\,\mathrm{d}a_m=\tilde{n}(m)\,\mathrm{d}m$. The quantity
$a_m^4n(a_m)$ is interpreted as the mass-weighted ($\propto a_m^3$)
grain size distribution per $\log a_m$. The grain size distribution is divided by
$n_\mathrm{H}$ to
cancel out the density difference between the ISM phases.
We refer to $a_m^4n(a_m)/n_\mathrm{H}$ as the grain size distribution as long as there is no
risk of confusion.

\subsection{Evolution of grain size distribution and filling factor}\label{subsec:gsd}

\begin{figure}
\includegraphics[width=0.45\textwidth]{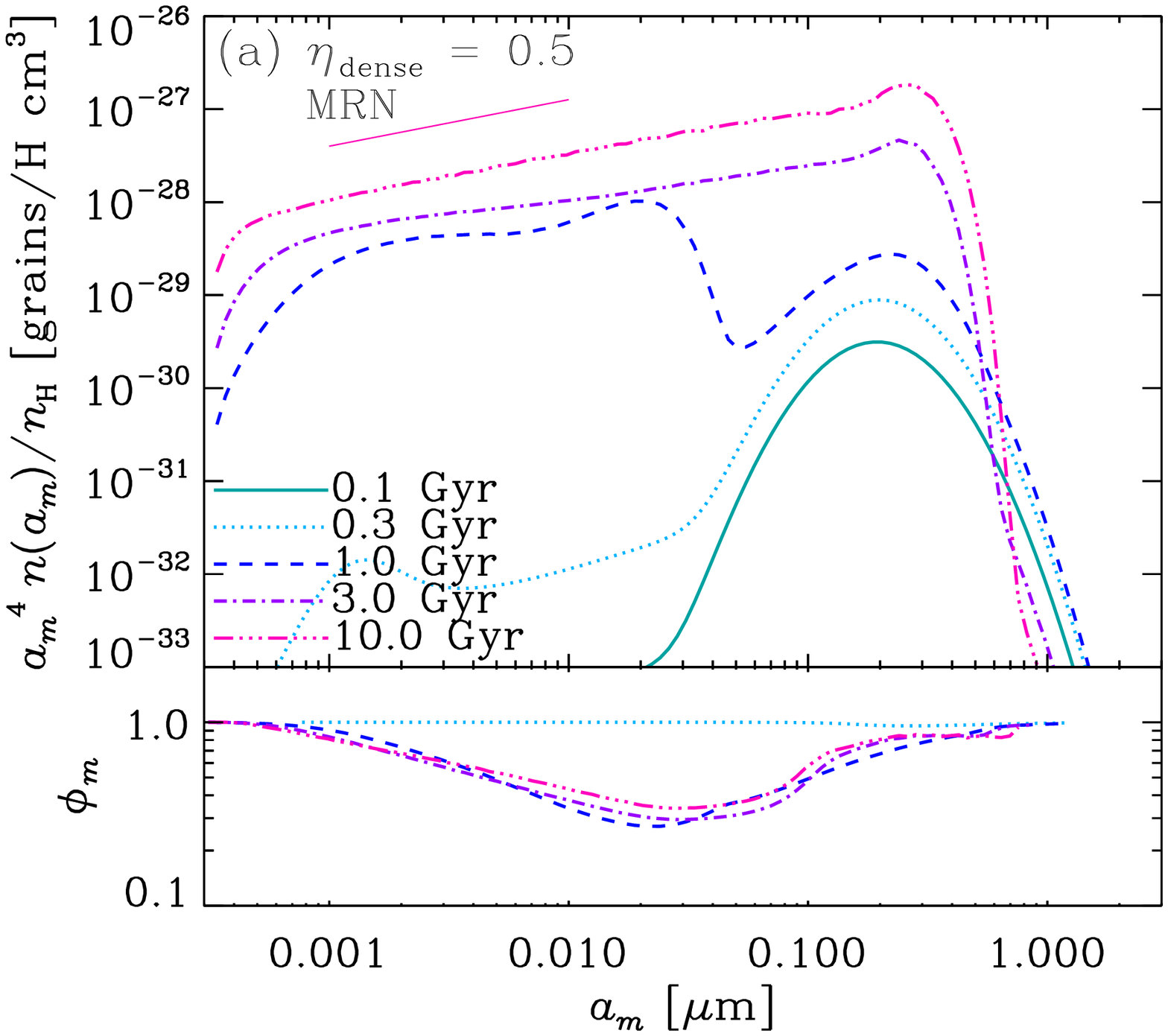}
\includegraphics[width=0.45\textwidth]{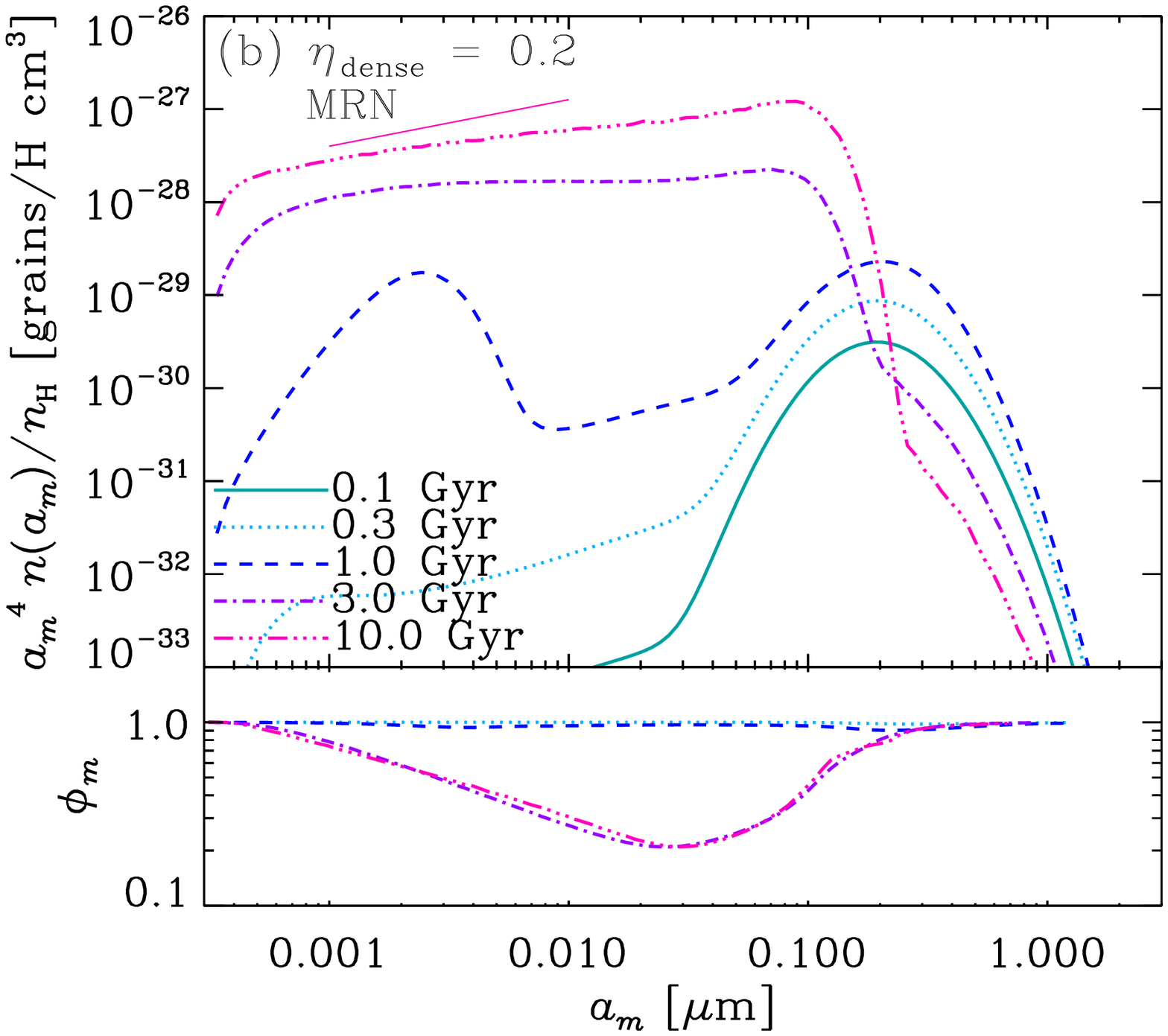}
\includegraphics[width=0.45\textwidth]{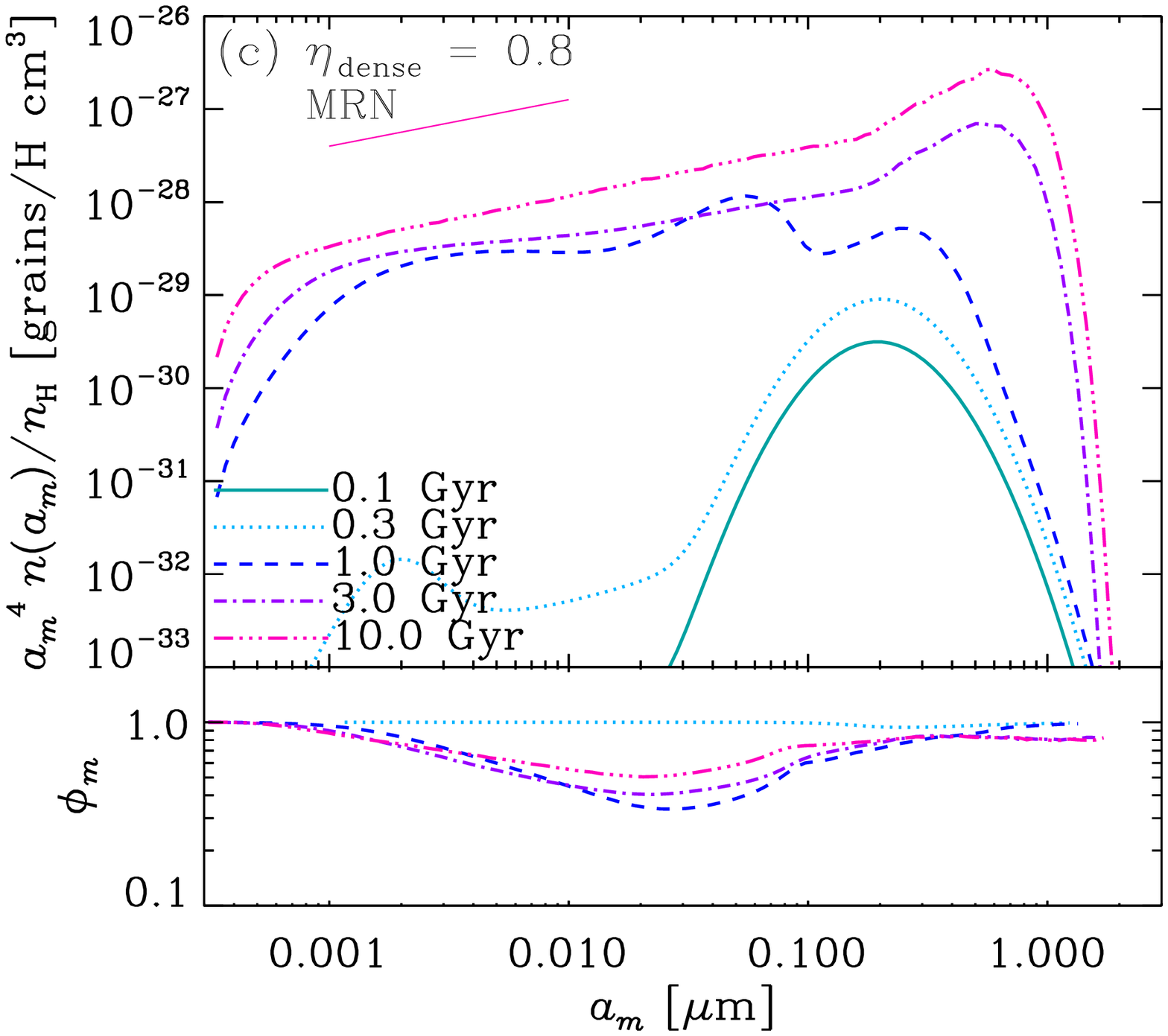}
\caption{Evolution of grain size distribution (upper window) and filling factor (lower window)
in each panel.
Panels (a), (b), and (c) show the results for $\eta_\mathrm{dense}=0.5$ (fiducial), 0.2, and
0.8, respectively.
The grain size distribution is multiplied by $a_m^4$
and divided by $n_\mathrm{H}$: the
resulting quantity is proportional to
the mass-weighted grain size distribution per $\log a_m$ relative to the gas mass.
The solid, dotted, dashed, dot--dashed, and triple-dot--dashed lines show
the results
at $t=0.1$, 0.3, 1, 3, and 10~Gyr, respectively.
For $\phi_m$, we do not present the result at $t=0.1$ Gyr, which is similar to that
at $t=0.3$ Gyr (i.e.\ unity for all $a_m$).
The thin straight solid line shows the MRN slope ($n\propto a_m^{-3.5}$).
\label{fig:gsd}}
\end{figure}

We present the evolution of grain size distribution and filling factor.
We show the results in Fig.~\ref{fig:gsd} for various $\eta_\mathrm{dense}$ with
a fixed $\tau_\mathrm{SF}=5$~Gyr.

First, we discuss the fiducial case ($\eta_\mathrm{dense}=0.5$; Fig.~\ref{fig:gsd}a).
The evolution of grain size distribution is overall similar to that presented in HA19.
In the early epoch ($t\lesssim 0.3$~Gyr), the grain abundance is
dominated by large ($a_m\sim 0.1~\micron$) grains from stellar sources.
At $t\sim 0.3$~Gyr, shattering gradually becomes prominent as we observe in the tail
of the grain size distribution at small $a_m$. Between $t=0.3$ and 1~Gyr, the small grain abundance
drastically increases because of dust growth by accretion. Note that accretion is more
efficient for smaller grains because of their larger surface-to-volume ratios.
After that, the abundance of large grains continuously grows because of coagulation.
As a consequence of continuous coagulation, the overall grain size distribution
approaches a power-law-like shape whose slope is similar to that derived by
\citet[][hereafter MRN]{Mathis:1977aa} for the MW extinction curve
($n\propto a_m^{-3.5}$). Some analytic studies suggested that
the slope of the grain size distribution reaches an
equilibrium and becomes similar to the MRN value
if the fragmentation and/or coagulation reach an equilibrium state
\citep[e.g.][]{Dohnanyi:1969aa,Tanaka:1996aa,Kobayashi:2010aa}.
%%\footnote{They mostly treated the fragmentation process (shattering),
%%but the same argument holds for coagulation;
%%for example, the analysis of equation (2.9) in \citet{Tanaka:1996aa}, which is valid for coagulation,
%%leads to the same equilibrium slope.}

The filling factor in the fiducial case decreases around $a_m\sim 0.03~\micron$
after $t\sim 1$~Gyr.
Since coagulation is the unique process that creates porosity {in our model},
the decrease of the filling factor coincident with the increase of large grains
is attributed to the onset of efficient coagulation.
The filling factor drops down to $\phi_m\sim 0.3$ around $a_m\sim 0.03~\micron$.
At $a_m\gtrsim 0.1~\micron$, compaction occurs
(recall that the grain velocity increases with $a_m$; equation~\ref{eq:vel}).
The filling factor stays almost unchanged after $t\sim 1$~Gyr;
in this `equilibrium' state, the porosity creation by coagulation is balanced
by the compact grain formation by shattering and accretion. This balance also realizes
the convergence of the slope in the grain size distribution.

Next, we discuss the effect of $\eta_\mathrm{dense}$.
We compare the results for $\eta_\mathrm{dense}=0.2$ and 0.8 with the fiducial case
in Fig.~\ref{fig:gsd}. In the early epochs ($t\lesssim 0.3$~Gyr),
the grain size distribution and the porosity are insensitive to $\eta_\mathrm{dense}$
because their evolution is dominated by stellar dust production, not by interstellar processing.
The evolution after $t\sim 1$~Gyr is, however, very different.

Here we discuss the case of $\eta_\mathrm{dense}=0.2$
(Fig.~\ref{fig:gsd}b). In this case, coagulation is inefficient
(compared with the fiducial case), which makes the following differences
in the grain size distribution at $t\gtrsim 1$~Gyr.
At $t=1$~Gyr, the small grain abundance is less for $\eta_\mathrm{dense}=0.2$ than for
$\eta_\mathrm{dense}=0.5$ because accretion is less efficient. After $t\sim 3$~Gyr,
the difference in the grain size distribution is explained by the inefficient coagulation,
which has the following
two influences: (i) The small-grain-dominated phase is more prominent because small
grains do not coagulate efficiently; and
(ii) the grain size distributions at later times ($t\sim 3$--10 Gyr) are more dominated by
small grains as seen in the slope (compared with the MRN distribution) and the maximum
grain radius. The filling factor is also affected by $\eta_\mathrm{dense}$ at later epochs.
Compared with the above fiducial case,  the filling factor drops more
(down to $\phi_m\sim 0.2$ at $a_m\sim 0.03~\micron$) or the porosity develops more.
This is because
more efficient shattering for smaller $\eta_\mathrm{dense}$ provides more small grains
from which coagulation creates porosity (H21).
The grain radius ($a_m$) at which $\phi_m$
becomes minimum is, however, insensitive to $\eta_\mathrm{dense}$. This radius corresponds to the
value above which compaction occurs.
Indeed, using equation (\ref{eq:vel}), the impact energy
is roughly estimated as $E_\mathrm{imp}\sim\frac{1}{2}mv_\mathrm{gr}^2\sim
3.2\times 10^{-10}(a_m/0.03~\micron )^4(\phi_m /0.3)^{2/3}(s/3.5~\mathrm{g~cm}^{-3})^2$ erg
%%3.21961e-10
in the dense ISM. The rolling energy is, on the other hand, approximately given by
$E_\mathrm{roll}\sim 8.9\times 10^{-10}(a_m/0.03~\micron )(\gamma /25~\mathrm{erg~cm}^{-2})
(\xi_\mathrm{crit}/10~\text{\AA})$ erg. Thus, $E_\mathrm{imp}\gtrsim E_\mathrm{roll}$
is satisfied if $a_m\gtrsim 0.04~\micron$, which explains the grain radius above which
compaction (increase of $\phi_m$) occurs. The filling factor also increases towards
small $a_m$ because shattering continuously supplies compact small grains.
Note that the above condition of compaction does not depend on $\eta_\mathrm{dense}$.

In the case of $\eta_\mathrm{dense}=0.8$, the maximum grain radius is larger than in the
fiducial case at
later epochs because of more efficient coagulation.
%%We also see a slight suppression of the tail towards smaller grain radii at $t\sim 0.1$ Gyr
%%because of inefficient shattering.
The filling factor is overall larger because the production of small grains, from which porosity
is created through coagulation, is less efficient.
The filling factor continues to increase up to $t\sim 10$ Gyr.
For large ($a\gtrsim 0.3~\micron$) grains, the porosity is determined
by the maximum compaction regulated by $\epsilon_V$ in Section \ref{subsubsec:coag};
that is, the grains are not completely compressed. This treatment has already been discussed by H21
and it does not affect our discussions below significantly.

%%Parameters: $E_\mathrm{roll}$, which regulates compaction. $E_\mathrm{roll}$ is proportional to
%%$\xi_\mathrm{crit}$).
%%The filling factor also converges to the same value determined roughly by
%%$1/(1+\epsilon_V)$ at the largest grain radii as mentioned above.
%%Grain compaction is also affected by the number of contacts $n_\mathrm{c}$
%%We also examine the dependence on $\epsilon_V$, which regulates the maximum compaction.

\subsection{Effect of star-formation time-scale}\label{subsec:tauSF}

{As shown by \citet{Asano:2013ab}, metal enrichment plays an important role in
determining the dust abundance, especially because dust growth by accretion is governed by
the metallicity.
Indeed, as argued by \citet{Hirashita:2020aa}, a similar grain size distribution
is realized at the same $t/\sqrt{\tau_\mathrm{SF}}$.
The stellar dust yield has a minor influence on the dust enrichment after
dust growth starts to dominate the total dust abundance.
Since the metal-enrichment time-scale is determined by $\tau_\mathrm{SF}$,
it is useful to examine the evolution of grain size distribution for various $\tau_\mathrm{SF}$.}

\begin{figure}
\includegraphics[width=0.45\textwidth]{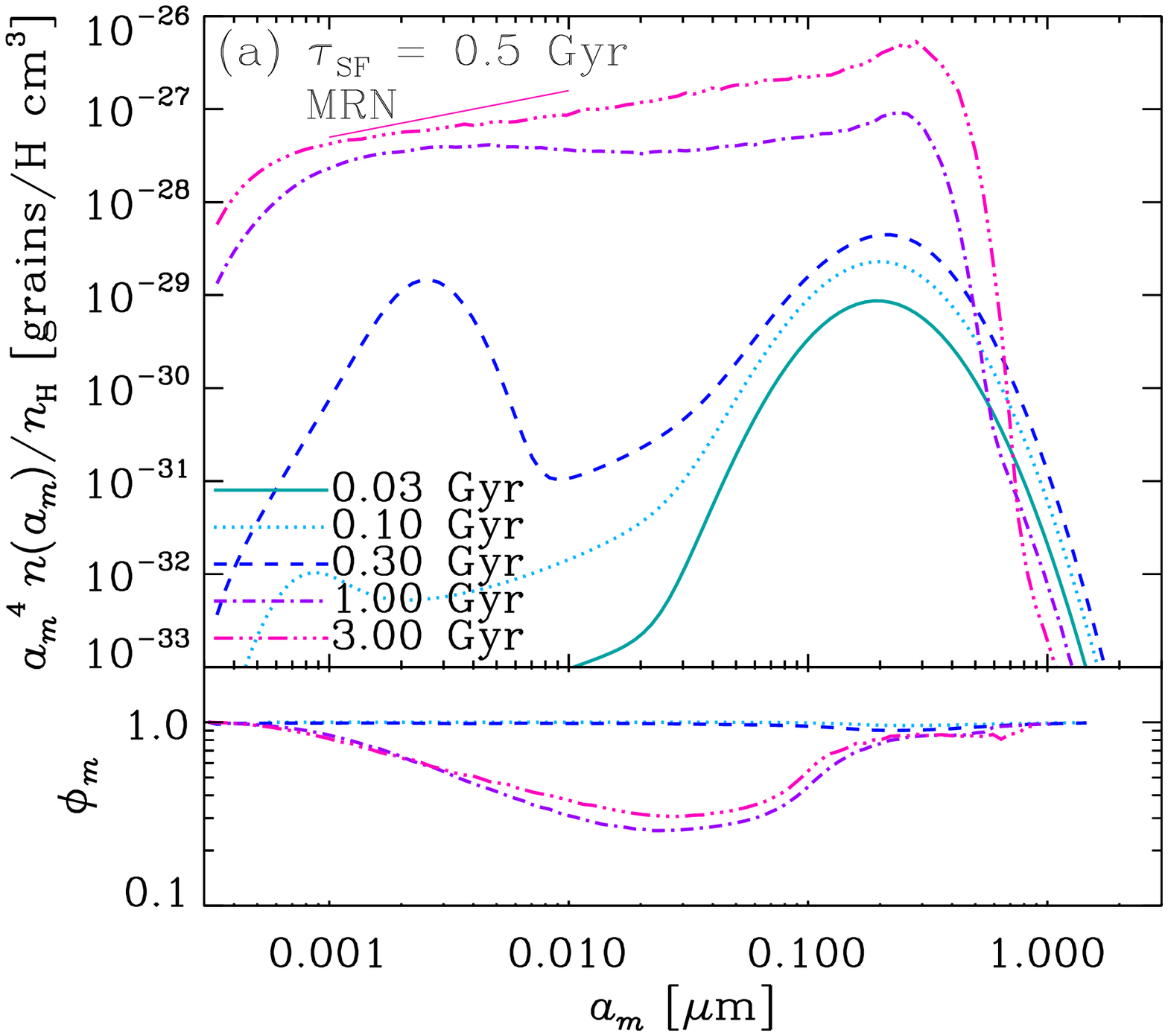}
\includegraphics[width=0.45\textwidth]{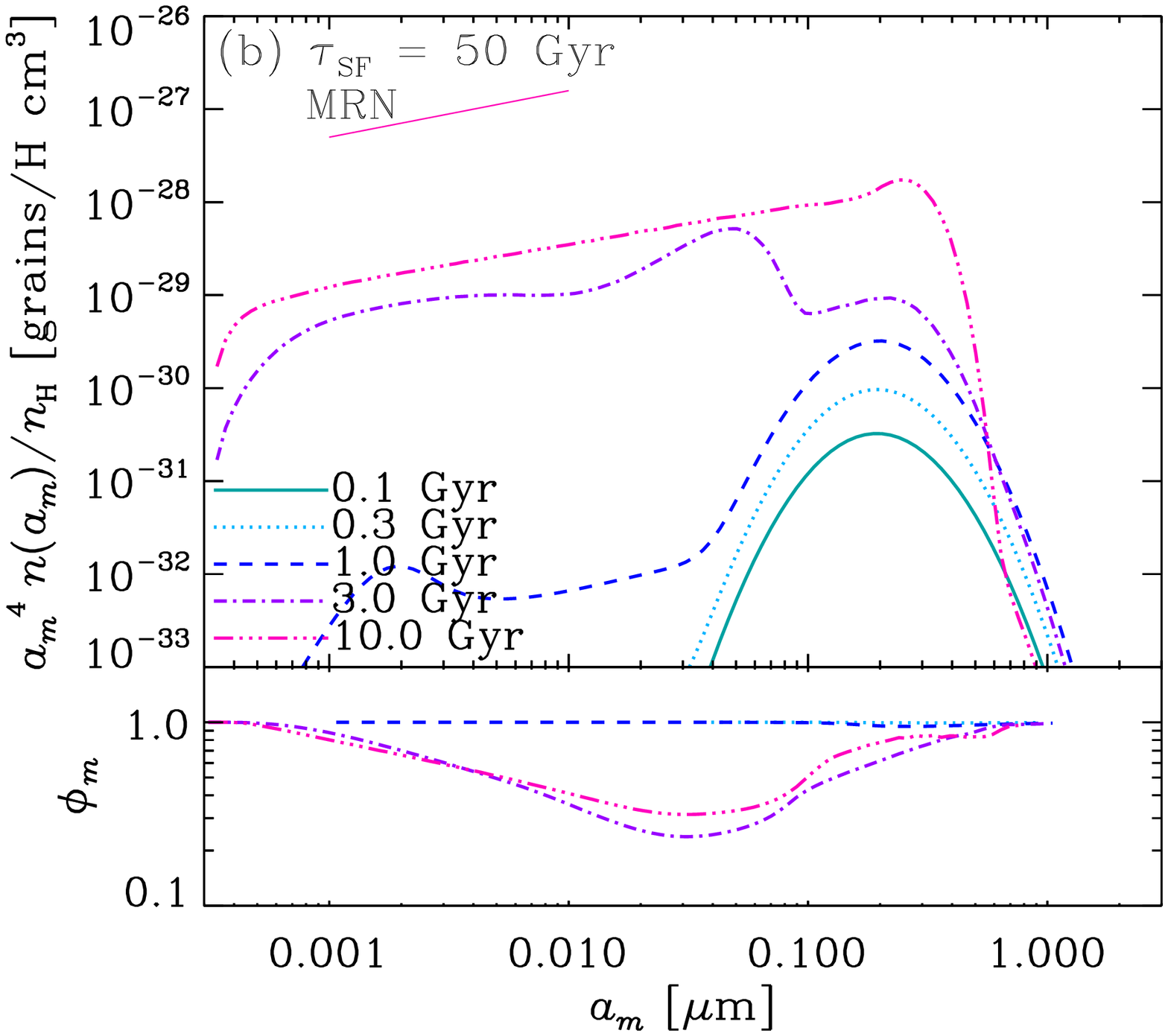}
\caption{Same as Fig.\ \ref{fig:gsd} but for various star formation time-scales $\tau_\mathrm{SF}$:
(a) $\tau_\mathrm{SF}=0.5$ Gyr and (b) 50 Gyr (with $\eta_\mathrm{dense}=0.5$).
The correspondence between the line species and the age is shown in the legend.
Note that the case with $\tau_\mathrm{SF}=5$ Gyr
(fiducial case) is shown in Fig.\ \ref{fig:gsd}a.
\label{fig:gsd_tauSF}}
\end{figure}

We present the evolution of grain size distribution and filling factor for various $\tau_\mathrm{SF}$
(0.5 and 50 Gyr) with fixed $\eta_\mathrm{dense}=0.5$
in Fig.\ \ref{fig:gsd_tauSF}. Note that the fiducial case with $\tau_\mathrm{SF}=5$ Gyr
is presented in Fig.\ \ref{fig:gsd}a.
For $\tau_\mathrm{SF}=0.5$ Gyr, since the metal enrichment occurs faster,
we also show $t=0.03$ Gyr
and stop at $t=3$ Gyr. In this case, except for the faster evolution, the overall evolutionary behaviour
of grain size distribution is similar to the results in the fiducial case. The filling factor and its
$a_m$ dependence are also similar between the two cases.
{As mentioned above, the evolutionary time-scale of grain size distribution
is scaled with $\propto\sqrt{\tau_\mathrm{SF}}$};
for example, a grain size distribution
smoothed by coagulation appears at $t\sim 1$ Gyr for $\tau_\mathrm{SF}=0.5$ Gyr while
it appears at $t\sim 3$ Gyr for $\tau_\mathrm{SF}=5$ Gyr. The porosity also develops when
coagulation starts to be efficient; thus, considering the above scaling, the
time when the effect of coagulation prominently appears is roughly evaluated as
$t\sim 3(\tau_\mathrm{SF}/5~\mathrm{Gyr})^{1/2}$ Gyr.

For $\tau_\mathrm{SF}=50$ Gyr, the dust enrichment proceeds more slowly and stops
at the lower dust abundance at $t=10$ Gyr than in the other cases with shorter $\tau_\mathrm{SF}$.
This is due to slow chemical enrichment. However, interstellar processing of dust still makes
the grain size distribution approach the MRN slope. The filling factor achieved is similar to the
above. Thus, the porosity, once it is created by coagulation, is insensitive to
$\tau_\mathrm{SF}$. Also, the grain radius at which the filling factor becomes minimum
does not depend on $\tau_\mathrm{SF}$ from the above argument in Section \ref{subsec:gsd}.
We can also confirm that the time-scale on which
the grain size distribution and the porosity are modified by interstellar
processing follows the
above scaling $\propto\tau_\mathrm{SF}^{1/2}$.

\subsection{Extinction curves}\label{subsec:ext_result}

\begin{figure}
\includegraphics[width=0.48\textwidth]{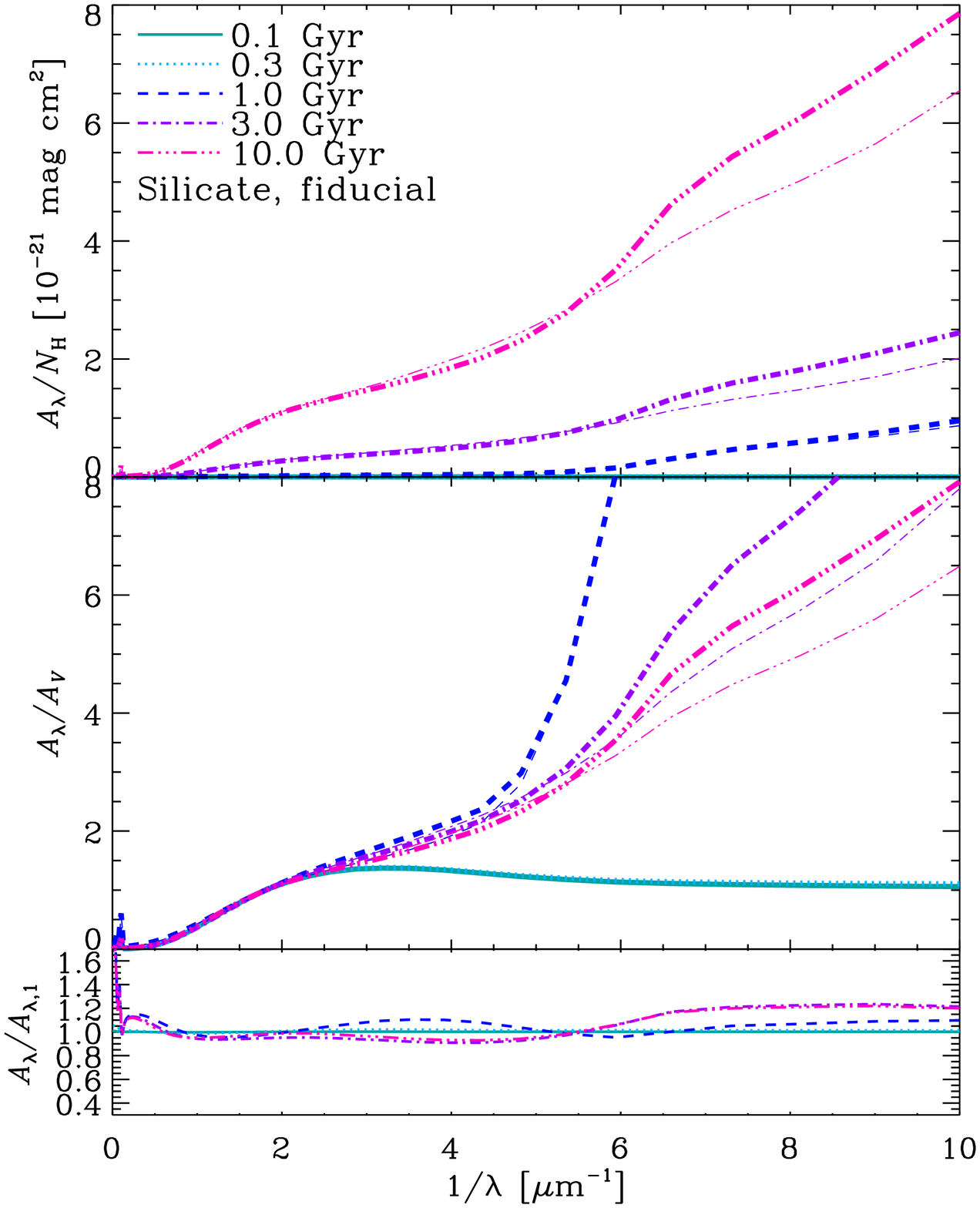}
\includegraphics[width=0.48\textwidth]{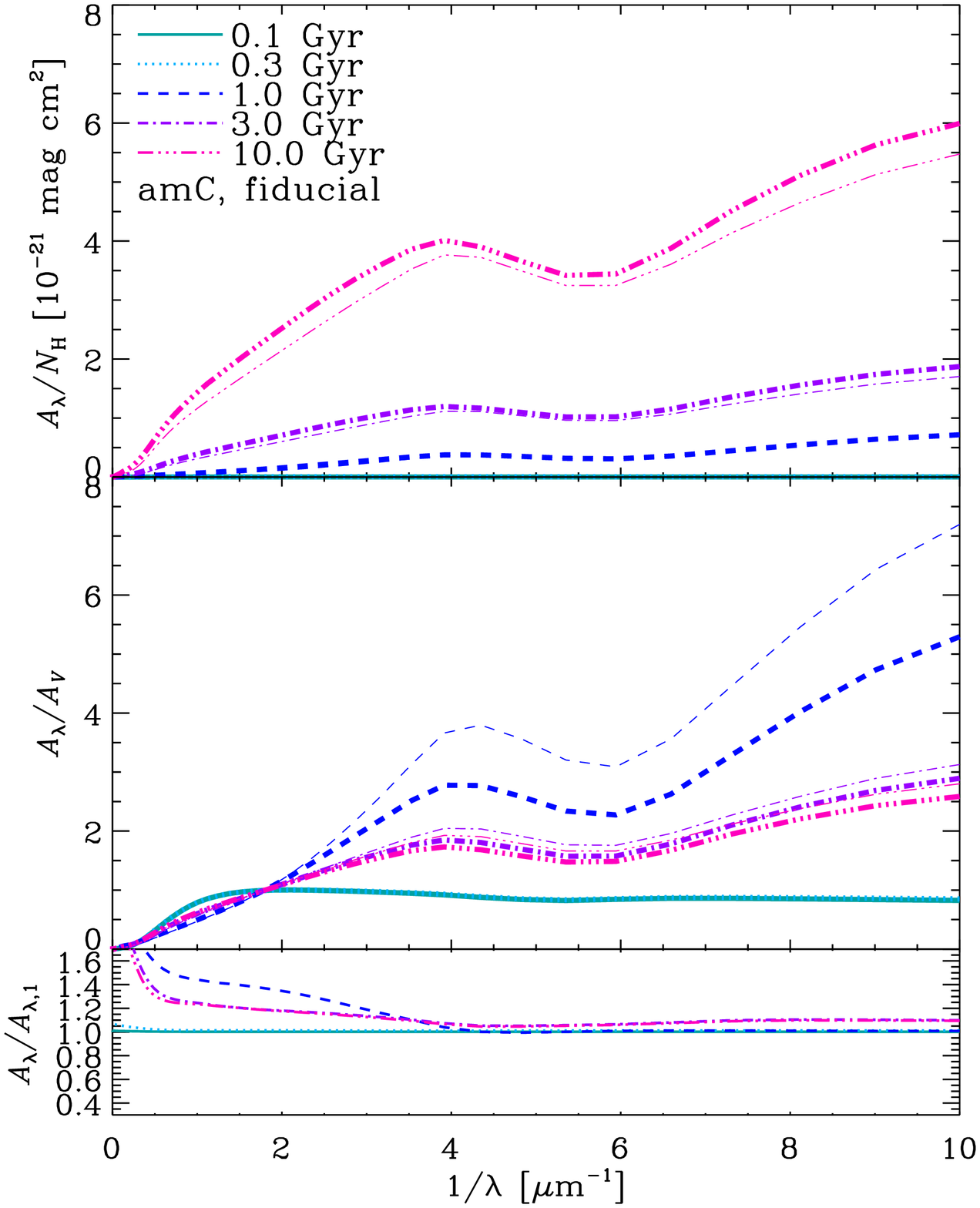}
\caption{Extinction curves for silicate and amC in the upper and lower panels, respectively.
The thick solid, dotted, dashed, dot--dashed, and triple-dot--dashed lines
show the extinction curves at $t=0.1$, 0.3, 1, 3, and 10 Gyr, respectively.
The thin lines show $A_{\lambda ,1}$ (extinction without porosity) with the same line species.
In each panel, the upper, middle, and lower windows present the extinction per hydrogen,
the extinction normalized to the $V$-band value, and the ratio of $A_\lambda$ to
$A_{\lambda ,1}$ (an indicator of the porosity effect), respectively.
\label{fig:ext}}
\end{figure}

We calculate the evolution of extinction curve (shown in two forms: $A_\lambda /N_\mathrm{H}$ and
$A_\lambda /A_V$) for silicate and amC separately based on the
grain size distributions and the filling factors presented above, using
the method in Section~\ref{subsec:ext_method}.
We also show the extinction curve without porosity, $A_{\lambda ,1}$,
defined in Section~\ref{subsec:ext_method} so that the effect of porosity can be quantified by
$A_\lambda /A_{\lambda ,1}$. We display the results in Fig.\ \ref{fig:ext}.

First, we examine the case of silicate.
From Fig.\ \ref{fig:ext} (upper panel), we observe that $A_\lambda /N_\mathrm{H}$ monotonically
rises at all wavelength because of dust enrichment. Compared with the extinction without
porosity ($A_{\lambda ,1}$), $A_\lambda /N_\mathrm{H}$ is higher in the far-UV and lower
in the near-UV at $t\geq 3$ Gyr, when the porosity has developed.
This is also clear in the ratio $A_\lambda /A_{\lambda ,1}$.
Thus, the porosity can both increase and decrease the extinction depending on the wavelength
as noted by \citet{Voshchinnikov:2006aa} and \citet{Shen:2008aa} (and discussed in H21).
The effect of dust enrichment is cancelled out if we show
$A_\lambda /A_V$, from which we observe that the extinction curve steepens
up to $t\sim 1$ Gyr and flattens after that. A similar evolutionary behaviour was also found by
our previous calculation without porosity (HA19), and is interpreted
by the evolution of grain size distribution shown in Section \ref{subsec:gsd}. The extinction curve is flat
in the beginning because the grain abundance is dominated by large grains (stellar dust production).
After that, the extinction curve becomes steep because of small grain production by shattering
and accretion.
The steepening at $t=1$ Gyr corresponds to the drastic
increase of grains with $a_m\lesssim 0.02~\micron$, which contribute to the extinction
at $\lambda\lesssim 2\upi a_\mathrm{ch}\sim 0.19~\micron$
($1/\lambda\gtrsim 5.3~\micron^{-1}$) if we adopt $\phi_m =0.3$
from Fig.\ \ref{fig:gsd} (recall that $a_\mathrm{ch}=a_m\phi_m^{-1/3}$).
This explains the steep rise for silicate around $1/\lambda\sim 5~\micron^{-1}$ at $t=1$ Gyr.
After $t=1$ Gyr, since coagulation efficiently converts small grains to large grains,
the extinction curve becomes flatter, but not as flat as the one in the early evolutionary stage.
At such later ages, the porosity makes the extinction curve shape ($A_\lambda /A_V$) steeper,
especially in the far-UV.
This is because of the above-mentioned behaviours of $A_\lambda /A_{\lambda , 1}$;
that is, the enhancement of far-UV extinction by porosity.

Next, we discuss the extinction curves of amC.
Fig.\ \ref{fig:ext} shows monotonic rise of $A_\lambda /N_\mathrm{H}$ as $t$ increases
because of dust enrichment
as also seen for silicate. The porosity created at $t\gtrsim 3$ Gyr enhances
$A_\lambda /N_\mathrm{H}$
at almost all wavelengths, which is clear also in $A_\lambda /A_{\lambda ,1}$.
Recall that the porosity does not necessarily enhance the extinction for silicate.
This difference between silicate and amC was already noted by H21.
The evolution of $A_\lambda /A_V$ for amC is qualitatively similar to that for silicate:
steepening up to $t\sim 1$ Gyr and subsequent flattening. However, the overall steepness is
less than in the case of silicate.
At $t\ge 1$ Gyr, if we compare $A_\lambda$ with $A_{\lambda ,1}$,
we find that the extinction curve shape is flattened by the porosity. This is because
the porosity enhances the extinction more around the $V$ band wavelength, where the
extinction curve is normalized, than in the UV.
Except at $t\sim 1$ Gyr, when the rapid
enhancement of the small grain abundance occurs,
porosity has only a moderate effect on the resulting extinction curve
shape of amC.

The porosity also enhances the opacity at IR wavelengths
($1/\lambda\ll 1~\micron^{-1}$) for both dust materials as also shown by previous studies
\citep[e.g.][]{Voshchinnikov:2006aa}. Naturally, the enhancement stays finite, so that
$A_\lambda$ always converges to zero as $1/\lambda$ approaches to zero.
As mentioned in the Introduction, we concentrate on the wavelengths shorter than near-IR
and do not discuss the behaviour at such
long wavelengths in this paper.

The porosity enhances the far-UV extinction by $\sim$20 per cent
for silicate and $\sim$10 per cent for amC at $t\gtrsim 3$ Gyr.
This means that we could
`save' 10--20 per cent of metals to realize a certain amount of
far-UV extinction.
%%This factor, in general, cannot be extremely large
%%because the increase of grain volume and the
%%`dilution' of permittivity have opposite effects on the extinction \citep{Li:2005aa}.
However, the porosity has an opposite effect for
the near-UV extinction of silicate, which is diminished by $\sim 10$ per cent.
Thus, it is not obvious if porosity really saves the metals.
We will further investigate this point in Section \ref{sec:MW}.

We also discuss the results with different $\eta_\mathrm{dense}$
(= 0.2 and 0.8) based on the grain size distributions shown in Fig.\ \ref{fig:gsd}.
Since the results are quite obvious,
we show the figures in Appendix \ref{app:ext_dependence}.

In Fig.\ \ref{fig:ext_eta02}, we present the evolution of extinction curve for
$\eta_\mathrm{dense}=0.2$.
The following discussions hold for both silicate and amC (unless the species is specified).
The difference from the fiducial case appears after $t\gtrsim 1$~Gyr, when
interstellar processing starts to dominate the grain size distribution.
At $t=1$~Gyr, the extinction curves in the case of $\eta_\mathrm{dense}=0.2$ are
rather flatter than those in the fiducial case because the increase of the small grain abundance by
accretion is slower (recall that accretion occurs in the dense ISM).
At later epochs ($t\gtrsim 3$~Gyr), the extinction curves are steeper than in the fiducial case
because of the dominance of small grains in the grain
size distribution (Fig.\ \ref{fig:gsd}b). For silicate, the steepness is further enhanced by the
effect of porosity, which is clear from $A_\lambda /A_{\lambda ,1}$: The extinction
at $1/\lambda\sim 1.5$--6 $\micron^{-1}$ is diminished, while that in the far-UV is
enhanced. This behaviour for silicate is already noted by \citet{Voshchinnikov:2006aa},
and the large porosity achieved in the case of $\eta_\mathrm{dense}=0.2$ makes the
effect of porosity prominent.
Because of inefficient coagulation in this case, the extinction curve shape ($A_\lambda /A_V$)
stays steep after $t=3$ Gyr for the case of silicate, contrary to the significant flattening in the case of
$\eta_\mathrm{dense}=0.5$. 

We also show the extinction curves for
the dense-gas-dominated case with $\eta_\mathrm{dense}=0.8$
in Fig.\ \ref{fig:ext_eta08}. The extinction curves are overall flatter compared with the
fiducial case for both silicate and amC.
This is because of the overall large grain radii (Fig.\ \ref{fig:gsd}c).
The flatness of the extinction curve compared with the other cases is confirmed in
the plots of $A_\lambda /A_V$ except for the steep phase at $t=1$ Gyr.
For the case of silicate, the flatness is not only due to
efficient coagulation but also because of less porosity (as discussed above,
porosity makes the silicate extinction curve slope at optical--UV wavelengths steeper).
However,
the extinction curves of silicate with porosity are still slightly steeper than those without porosity
at $t\ge 3$ Gyr.
The porosity effect is less prominent in amC.

Finally, we discuss the effect of $\tau_\mathrm{SF}$. We calculate the extinction curves
for $\tau_\mathrm{SF}=0.5$ and 50~Gyr
based on the grain size distributions presented in Fig.~\ref{fig:gsd_tauSF}.
We show the results for $\tau_\mathrm{SF}=0.5$ and 50 Gyr in Appendix \ref{app:ext_dependence}
(Figs.\ \ref{fig:ext_tau05} and \ref{fig:ext_tau50}, respectively).
As mentioned above, $\tau_\mathrm{SF}$ regulates the overall time-scale of dust enrichment.
Thus, the evolution of extinction curve occurs
on a different time-scale depending on $\tau_\mathrm{SF}$.
%%although it follows the same evolutionary trend as in the fiducial case.
In particular, the level of
$A_\lambda /N_\mathrm{H}$ reflects the dust abundance, so that it rises more quickly
for shorter $\tau_\mathrm{SF}$. As mentioned above, the time-scale of interstellar processing is
scaled as $\tau_\mathrm{SF}^{1/2}$.
For example, in the case of $\tau_\mathrm{SF}=0.5$ Gyr
(50 Gyr), the steepening of extinction curve starts around $t= 0.3$ Gyr (3 Gyr), while a similar
steepening appears
around $t=1$ Gyr in the fiducial case with $\tau_\mathrm{SF}=5$ Gyr. Thus, the steepening of
the extinction curve happens around $t\sim (\tau_\mathrm{SF}/5~\mathrm{Gyr})^{1/2}$ Gyr,
considering the above scaling, although the extinction curves are not exactly the same
because of
the different scaling of the chemical enrichment time-scale ($\propto\tau_\mathrm{SF}$).

\section{The MW extinction curve}\label{sec:MW}

Although the main purpose of this paper is to give basic predictions on the
porosity formation in galaxy evolution, it may be useful to discuss how the
above results fit the actually observed extinction curves.
Here, we focus on the most investigated target -- the MW extinction curve.
The purpose of this section is to examine if the predicted porosity is
still accepted by the observed MW extinction curve. We do not aim at
detailed fitting to the MW extinction curve.
This section will provide a key for further making a detailed model of the MW dust
based on our dust evolution calculations.

There is still clearly a missing component in our model to explain the MW extinction curve --
the 2175 \AA\ bump carriers.
As mentioned in Section \ref{subsec:ext_method}, if graphite has porosity,
the central wavelength of the bump changes,
which is not observed in the MW.
Our interpretation is that the bump is attributed to a component
separated from the general porosity evolution.
This implies that the modelling of the 2175 \AA\ bump carrier needs a special treatment.
We propose a possibility that polycyclic aromatic hydrocarbons
(PAHs) are responsible for
the 2175 \AA\ bump \citep{Li:2001aa}, and may not be suitable for being modelled using the
bulk properties as done in this paper.
In this section, we use the PAH component, of which the abundance is adjusted to fit
the 2175 \AA\ bump, but
we adopt the above calculation results for other grain materials (silicate and amC).

We use $A_\lambda /N_\mathrm{H}$ for the extinction curve, since it includes
the information on the dust abundance. We use the extinction curves in the fiducial case
($\eta_\mathrm{dense}=0.5$
and $\tau_\mathrm{SF}=5$~Gyr) at $t=10$ Gyr (comparable to the age of the
MW).
%%Given that the extinction curve shapes are similar between $t=3$ and 10 Gyr,
%%using the results at $t=10$ Gyr does not change the conclusions significantly.
The above extinction curves are normalized by the dust-to-gas ratio (equation \ref{eq:dg})
and obtain $A_\lambda /N_\mathrm{H}/\mathcal{D}$ for silicate and amC.
We also calculate $A_\lambda /N_\mathrm{H}/\mathcal{D}$ for the PAH component in the
following way.
We take the absorption cross-section of PAHs per carbon atom, which is denoted as
$C_\mathrm{abs,C}$, from \citet{Li:2001aa}. For the PAH component, the dust-to-gas ratio,
which is defined as the PAH abundance per gas mass is
evaluated as $\mathcal{D}=12N_\mathrm{C}/(\mu N_\mathrm{H}$), where $N_\mathrm{C}$ is
the column density of carbon atoms (we neglect the contribution of hydrogen to
the PAH mass because it only changes the mass by 4 per cent even if we assume H/C = 0.5).
Since the extinction is evaluated as
$A_\lambda =(2.5\log\mathrm{e})C_\mathrm{abs,C}N_\mathrm{C}$, we
obtain $A_\lambda/N_\mathrm{H}/\mathcal{D}=(2.5\log\mathrm{e})\mu C_\mathrm{abs,C}/12$
(recall that $\mu =1.4$) for PAHs.

In the fitting, we use $A_\lambda /N_\mathrm{H}/\mathcal{D}$ of each component from
the model, and we derive the abundance of each dust component from the fitting.
(In other words, we do not use the dust abundance calculated in the model.)
The total extinction per hydrogen is now written as
\begin{align}
\frac{A_\lambda}{N_\mathrm{H}} &=
\mathcal{D}_\mathrm{sil}(A_\lambda/N_\mathrm{H}/\mathcal{D})_\mathrm{sil}+
\mathcal{D}_\mathrm{amC}(A_\lambda/N_\mathrm{H}/\mathcal{D})_\mathrm{amC}\nonumber\\
&+
\mathcal{D}_\mathrm{PAH}(A_\lambda/N_\mathrm{H}/\mathcal{D})_\mathrm{PAH},
\end{align}
where $(A_\lambda/N_\mathrm{H}/\mathcal{D})$ for each component is distinguished by the
subscript (`sil', `amC', and `PAH' for the silicate, amC, and PAH components, respectively), and
$\mathcal{D}$ is the dust-to-gas ratio (the mass abundance relative to the gas mass)
of each component (the component is
indicated by the subscript).

\begin{table}
\caption{MW extinction curve fitting.}
\begin{center}
\begin{tabular}{lcccc}
\hline
Model & $\mathcal{D}_\mathrm{sil}$ & $\mathcal{D}_\mathrm{amC}$ &
$\mathcal{D}_\mathrm{PAH}$ & $\mathcal{D}_\mathrm{tot}$ \\
 & ($10^{-3}$) & ($10^{-3}$) & ($10^{-3}$) & ($10^{-3}$)\\
\hline
Fiiducial & 2.2 & 2.0 & 0.65 & 4.8 \\
$\eta_\mathrm{dense}=0.2$ & 0.34 & 2.2 & 0.43 & 3.0\\
$\eta_\mathrm{dense}=0.8$ & 9.8 & 0.67 & 0.85 & 11\\
w/o porosity$^a$ & 3.2 & 2.1 & 0.55 & 5.8 \\
Larger porosity$^b$ & 2.1 & 2.1 & 0.65 & 4.9 \\
\hline
\end{tabular}
\label{tab:fitting}
\end{center}
$^a$Using the fiducial result but forcing the porosity to be unity.\\
$^b$Case with enhanced porosity examined in Section \ref{subsec:discuss_porosity}.
\end{table}

\begin{figure}
\begin{center}
\includegraphics[width=0.45\textwidth]{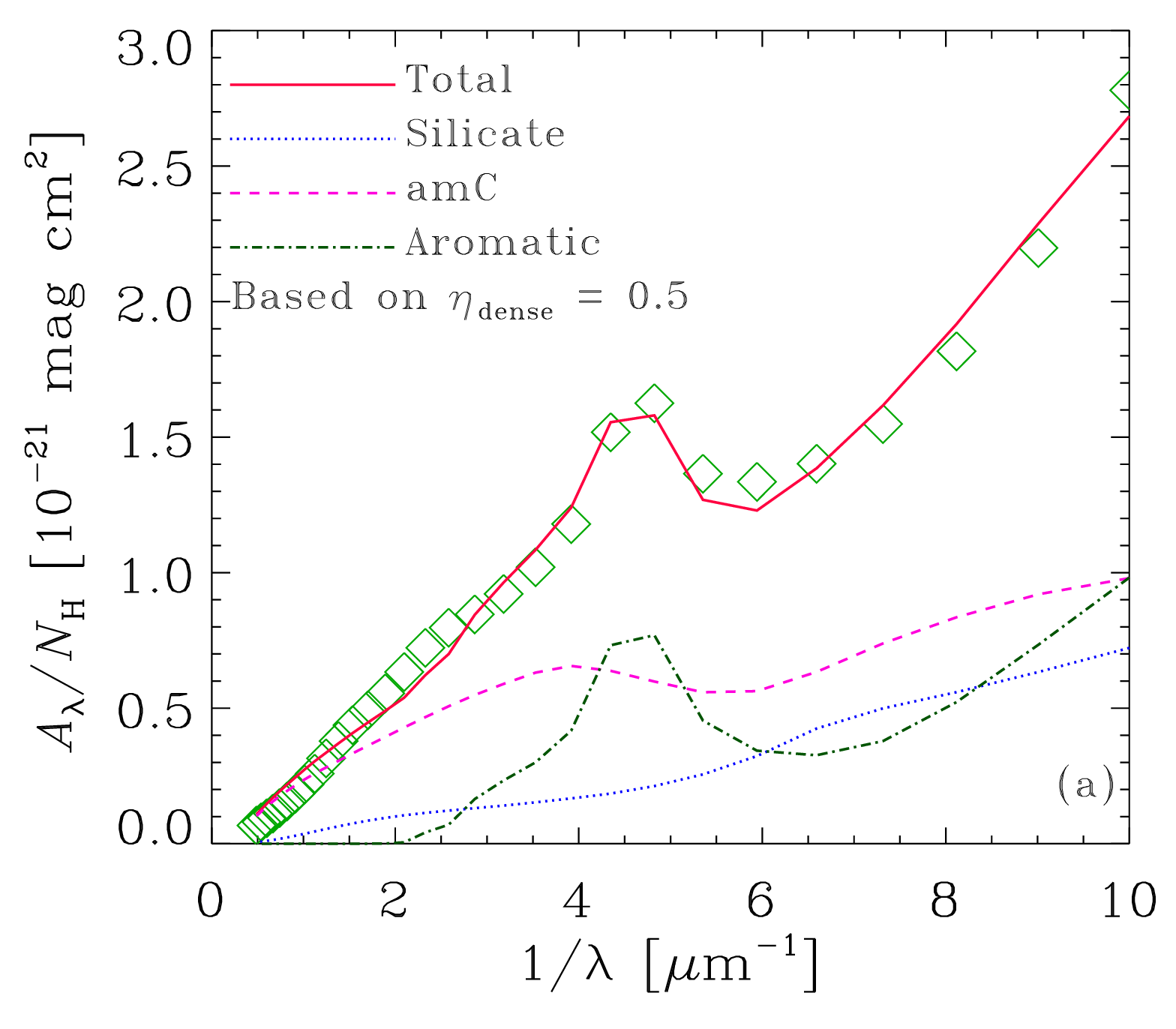}
\includegraphics[width=0.45\textwidth]{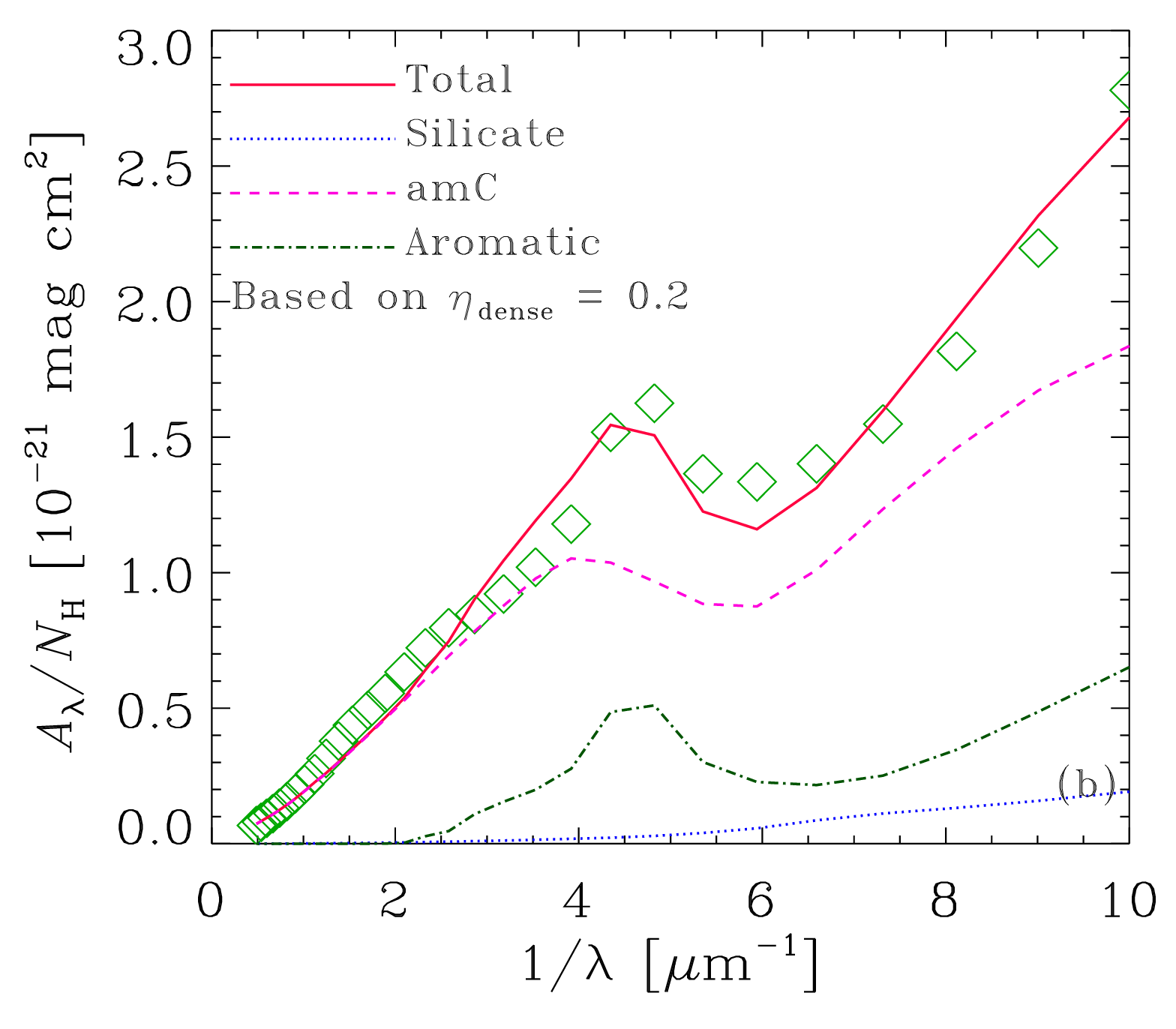}
\includegraphics[width=0.45\textwidth]{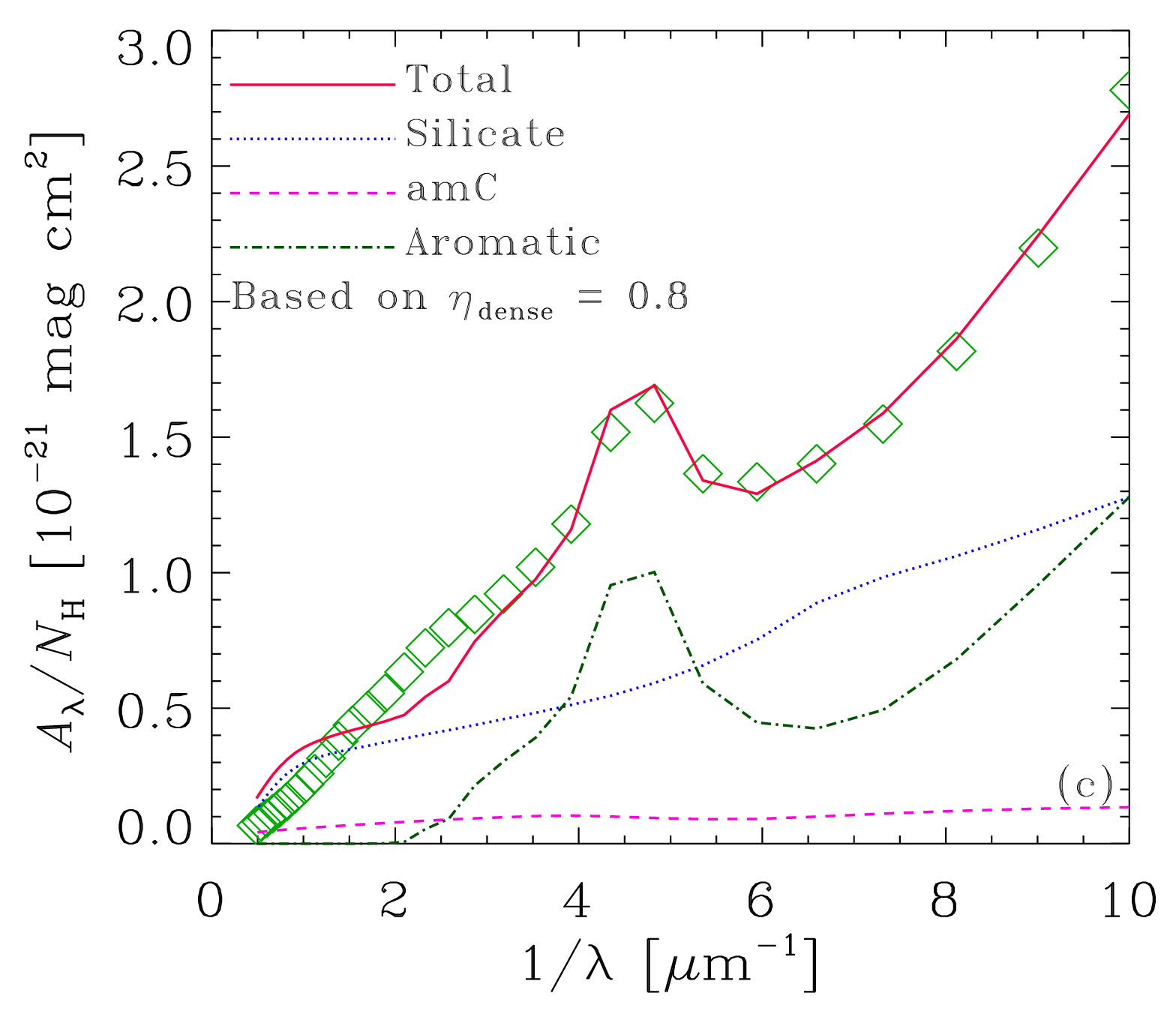}
\end{center}
\caption{Fitting to the MW extinction curve. The extinction curves
for the silicate and amC components are based on the models of
$\eta_\mathrm{dense}=0.5$ (fiducial), 0.2, and 0.8
($\tau_\mathrm{SF}=5$~Gyr and $t=10$ Gyr for all the models)
in Panels (a), (b), and (c), respectively.
We also add the PAH component. The abundances of the three dust components
are adjusted to fit the observational data points.
The solid line shows the best-fit total extinction, and the dotted, dashed, and dot--dashed lines
present the contribution from the silicate, amC, and PAH components, respectively.
The diamonds show the MW extinction curve adopted for the fitting.
\label{fig:fitting}}
\end{figure}

We adopt $A_\lambda /N_\mathrm{H}$ for the Milky Way extinction curve from
\citet{Cardelli:1989aa} [for $R_V=3.1$, where $R_V\equiv A_V/(A_B-A_V)$]
with the normalization determined by $A_V/N_\mathrm{H}=5.3\times 10^{-22}$ mag cm$^2$
\citep{Weingartner:2001aa}. We derive the values of $\mathcal{D}$ for the three components
by performing least square fitting. We list the best-fitting values of $\mathcal{D}$ in
Table~\ref{tab:fitting},
and show the fitting results in Fig.~\ref{fig:fitting}. We sample the wavelengths with a logarithmically
equal spacing from 0.1 to 2 $\micron$ (the data points we use are shown in Fig.\ \ref{fig:fitting}).

From Fig.~\ref{fig:fitting}a, we observe that the MW extinction curve is broadly reproduced by
using the fiducial model: At optical--near-IR wavelengths, the extinction level is determined
by amC,
since the other two components are relatively transparent. As expected, the PAH component
dominates the extinction around the 2175~\AA\ bump. The steep far-UV rise is
further supplemented by silicate and PAHs.
In this fitting, the abundances of silicate and amC are comparable
as shown in Table \ref{tab:fitting}. The total dust-to-gas ratio is
$4.8\times 10^{-3}$, slightly lower than the often used value ($\sim 6\times 10^{-3}$;
e.g.\ \citealt{Weingartner:2001aa}).
{However, it is not fair to compare these values directly since the goodness of
fitting is different.
%%As discussed later, the dominant reason for the difference
%%is not the porosity but the adopted grain size distribution.
It is natural that our fitting to the MW extinction
curve is worse than previous papers (because we do not change the grain size distribution
freely); nevertheless, our results still fit the MW extinction curve quite successfully.} The
slight overprediction in the near-IR and underproduction in the optical are
worth improving in the future. 

%%the abundance constraint for carbon is also severe (at most
%%$\mathcal{D}_\mathrm{amC}+\mathcal{D}_\mathrm{PAH}=2.0\times 10^{-3}$;
%%\citealt{Weingartner:2001aa}). Although there is an uncertainty in this upper limit, we
%%use slightly larger amount of carbon ($2.7\times 10^{-3}$).

\begin{figure}
\begin{center}
\includegraphics[width=0.45\textwidth]{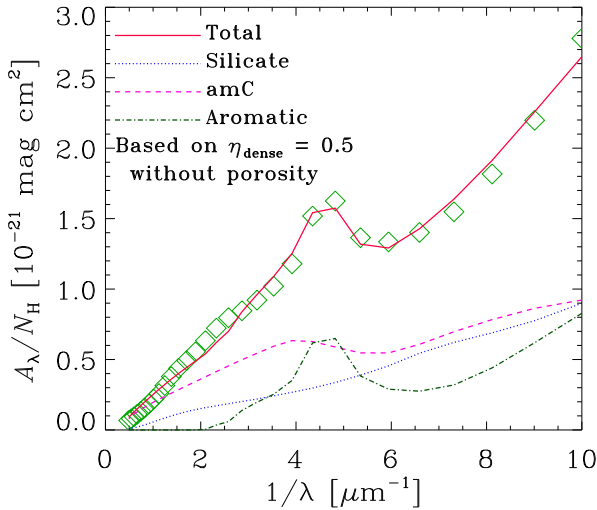}
\end{center}
\caption{Same as Fig.\ \ref{fig:fitting} (only for $\eta_\mathrm{dense}=0.5$)
but the fitting is performed using the extinction curves without porosity
(i.e.\ using $A_{\lambda ,1}$ for the extinction curves of silicate and amC).
\label{fig:fitting_wo_porosity}}
\end{figure}

To clarify the effect of porosity, we also perform another fitting using the extinction curves
without porosity ($\phi_m=1$; i.e.\ using $A_{\lambda, 1}$).
We show the fitting result in Fig.\ \ref{fig:fitting_wo_porosity}, and the resulting abundance
of each component in Table \ref{tab:fitting}. We observe that the best fit solution in this case has a
1.5 times higher silicate abundance and shows a
slightly better fit to the near-IR extinction than that in the above fit.
This is because the less steep extinction curve
of compact silicate grains fits the overall steepness of the MW extinction curve.
Nevertheless,
we also need slightly more amC in this case because compact amC has less extinction per dust mass
(see Fig.\ \ref{fig:ext}). However, less PAHs are required because of higher contribution of silicate to
the overall UV extinction. Thus, in total, the carbon abundance required for the MW fitting
is almost the same between the porous (fiducial) and non-porous cases.
The fact that the resulting extinction curves for the porous and non-porous
cases are similar means that the porosity is not the main driver to regulate the goodness of
fitting. {Note, though, that the resulting best-fit silicate abundance is affected by the porosity.}

We also examine the fitting using the results with different $\eta_\mathrm{dense}$.
If we use the grain size distribution for $\eta_\mathrm{dense}=0.2$ (with the other
parameters same as the fiducial case), we obtain the fitting result shown in
Fig.\ \ref{fig:fitting}b. As shown in Table \ref{tab:fitting}, the resulting abundance of silicate
is particularly small. This is mainly because of the difference in the grain size distribution.
The extinction curve of amC is steep enough to reproduce the overall slope of the
MW extinction curve. This leads to an extinction curve dominated by amC.
Silicate has too steep an extinction curve in this case, so that it is disfavoured.
The large porosity in this case also contributes to steepening the silicate extinction curve.
As a result, the silicate abundance in the best fit solution is only $\sim 3\times 10^{-4}$,
which is too small to explain the interstellar depletion \citep{Weingartner:2001aa}.
The fitting with $\eta_\mathrm{dense}=0.8$ provides an opposite extreme as shown in
Fig.\ \ref{fig:fitting}c.
In this case, the fitting solution is dominated by silicate, since the amC extinction curve is too
flat. The required silicate abundance is $\sim 0.01$
(Table \ref{tab:fitting}), which exceeds the
elemental abundance constraint \citep{Weingartner:2001aa}.
Moreover, the fitting in the near-IR--optical is bad; the extinction curve
at these wavelengths is too flat because of the dominance of large grains (Fig.\ \ref{fig:gsd}c).

It is interesting to point out that the fiducial case provides the best fit solution for
the MW extinction curve with a balanced grain abundance.
We also find that the grain size distribution predominantly determines the
goodness of the fit, and the porosity, as far as it is around our predicted value,
does not affect the shape of the best-fit extinction curve significantly. Thus, extinction curves
do not provide a strong constraint on the porosity, but rather constrain the grain
size distribution. In other words, the moderate grain porosities predicted in our paper
do not alter the overall understanding of extinction curves obtained from fitting efforts
in previous studies (e.g.\ MRN; \citealt{Kim:1994aa,Weingartner:2001aa,Zubko:2004aa}).

\section{Discussion}\label{sec:discussion}

\subsection{Uncertainties}\label{subsec:uncertainties}

The grain-size-dependent porosity is the prediction made possible in our
new framework that treats the porosity and the grain size distribution {simultaneously}.
As shown above (and in H21), the interplay between coagulation and shattering is
important for the porosity. However, the grain radius at which the porosity peaks
as well as the peak porosity value still depends on some parameters in coagulation and shattering.
We discuss which parameter could change the porosity evolution.

For coagulation, the results depends on
$\gamma$, $\xi_\mathrm{crit}$ and $\epsilon_V$ ($n_\mathrm{c}$ has a smaller
influence than $\xi_\mathrm{crit}$; H21).
Note that $\gamma$ is degenerate with $\xi_\mathrm{crit}$.
A small value of $\gamma$ or $\xi_\mathrm{crit}$ (equivalent to small $E_\mathrm{roll}$)
leads to small porosity, especially at large
($a_m\gtrsim 0.1~\micron$) radii, because compaction easily occurs. The parameter
$\epsilon_V$, which determines the maximum compaction, affects the porosity at
large grain radii, where strong compaction occurs ($a_m\gtrsim 0.3~\micron$).
Since such large grains are also processed by shattering, $\epsilon_V$ is also degenerate
with the treatment of compaction in shattered remnants. Also, shattering effectively
disrupt large grains at $a_m\gtrsim 0.3~\micron$. Thus, $\epsilon_V$ is
less important than $\gamma$ and $\xi_\mathrm{crit}$.
In summary, among the parameters that regulate coagulation,
$\xi_\mathrm{crit}$ and $\gamma$, which directly affect $E_\mathrm{roll}$, most efficiently
influence the porosity in our model.

The treatment of shattering also has some freedom. The degree of compaction
of shattered remnants may affect the porosity evolution. However, this only has a minor
influence on the porosity unless the porosity is very large (such as $\phi_m\lesssim 0.1$; H21).
As shown above, shattering has a positive effect on the porosity by supplying
fragments from which coagulation builds up porous grains. Thus, the efficiency of
fragment formation (i.e.\ the shattered fraction in a grain--grain collision)
is important for the porosity.
The shattered fraction is determined by the ratio between the specific impact energy and
$Q_\mathrm{D}^\star$; thus, both grain velocities and $Q_\mathrm{D}^\star$
affect the resulting grain porosity.
If we adopt a smaller value of
$Q_\mathrm{D}^\star$ (or larger grain velocities) than adopted above,
shattering becomes more efficient, leading to a result similar to
lower $\eta_\mathrm{dense}$; that is, the grain size distribution is more biased towards
smaller radii and the porosity becomes larger.
%%(as mentioned also in Section \ref{subsec:improve}).
In this sense, $\eta_\mathrm{dense}$ and
$Q_\mathrm{D}^\star$ are degenerate, and a case with
smaller $Q_\mathrm{D}^\star$ (or larger grain velocities) can be effectively investigated
by adopting smaller $\eta_\mathrm{dense}$.

%%The other processes, SN destruction and accretion, basically conserve the total number of the grains
%%(Section \ref{subsec:dest_acc}). In contrast to shattering, which increases the total
%%number of grains (ingredients of porous grains), SN destruction and accretion conserve
%%the total number of grains, and thus, cannot supply a large number of grains to build up porous grains.
%%Moreover, these two processes do not create porosity by themselves.
{The parameters related to the other processes (stellar dust production, SN destruction and accretion)
have minor effects on the porosity compared with those of shattering and coagulation.
This, however, does not mean that these processes are unimportant: They play important roles in
the evolution of grains size distribution and dust abundance (H19) and indirectly influence the efficiency
of coagulation and shattering. In particular, accretion drastically increases the abundance of small grains,
which are later coagulated.
Moreover, it is also interesting to point out that the
`bombardments' associated with sputtering and shattering could imprint some holes in dust
grains, which can be taken as a creation of porosity. Thus, in the future, it is still worth including
some other ways of creating grain inhomogeneity that could lead to the formation of porosity.}

To summarize, {in our model,}
the parameters concerning shattering and coagulation potentially affect the
results significantly.
Since the effects of these processes on the porosity have already been investigated in H21,
we do not repeat the parameter surveys. Nevertheless, since we newly included all processes, it
is worth reexamining some parameter dependence in the context of our new framework.
We discuss the effect of porosity creation further and implications for the MW extinction
curve in the following subsection.

\subsection{Effects of porosity on the evolution of grain size distribution}\label{subsec:discuss_porosity}

As discussed in the previous subsection, the parameters concerning shattering
and coagulation are important in regulating the porosity; in particular,
$\gamma$ and $\xi_\mathrm{crit}$
directly regulate the rolling energy $E_\mathrm{roll}$ (important for compaction), and
the ratio between the specific impact energy and $Q_\mathrm{D}^\star$
governs the efficiency of producing shattered fragments (from which porous grains
form through coagulation).
In order to examine how much the effects of porosity could be enhanced,
we focus on the cases where porosity increases compared with the above results.

First, we consider a case with decreased $Q_\mathrm{D}^\star$, so that shattering becomes more efficient
than in the fiducial case. As mentioned in the previous subsection, the result with smaller $Q_\mathrm{D}^\star$
becomes similar to that with smaller $\eta_\mathrm{dense}$.
Thus, the fitting to the MW extinction curve becomes similar to the case shown
in Fig.\ \ref{fig:fitting}b (Section \ref{sec:MW}). In this fitting solution, silicate is
not favoured, leading to too small a silicate abundance.
This is not supported by the interstellar depletion as discussed above.

\begin{figure}
\begin{center}
\includegraphics[width=0.46\textwidth]{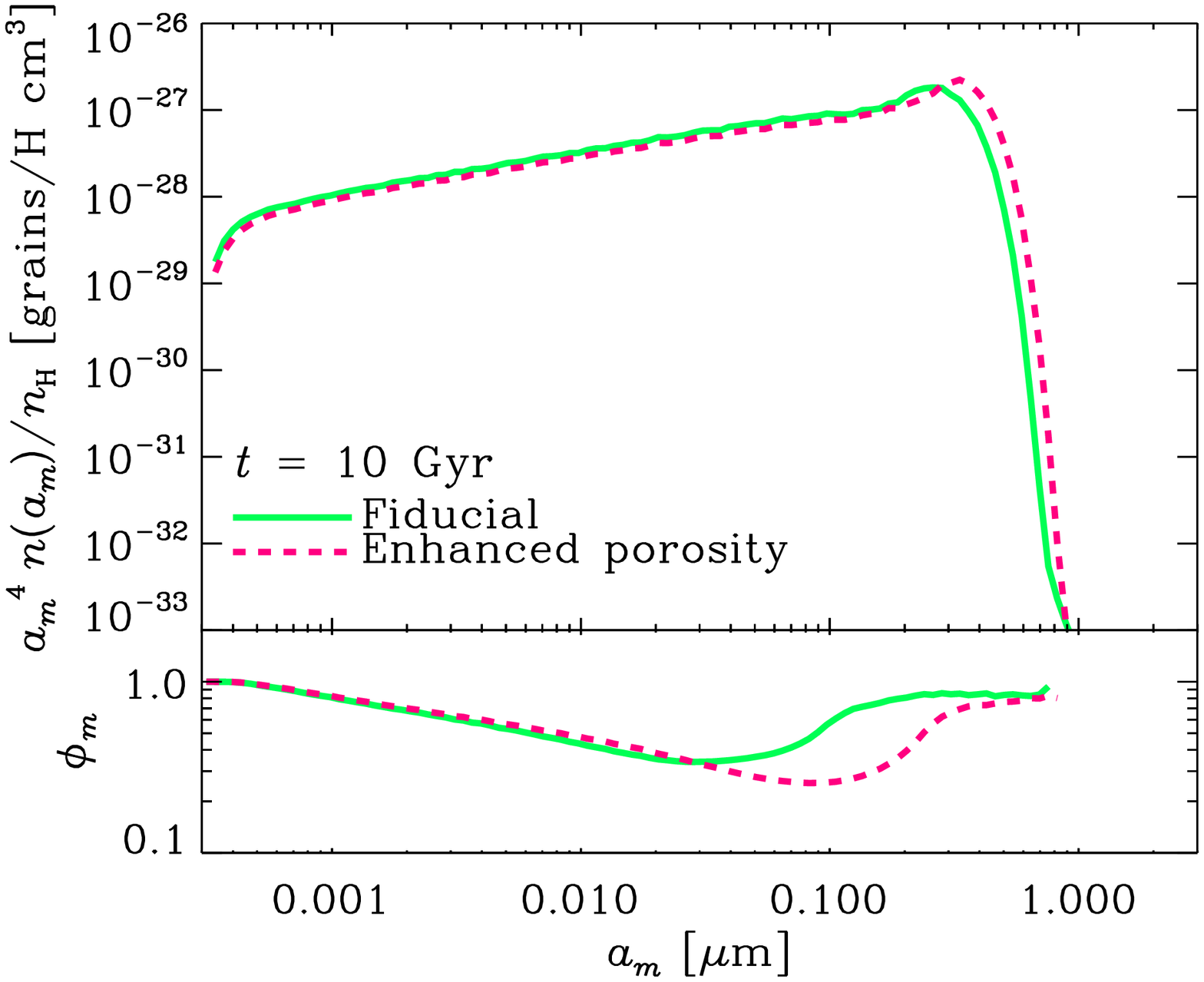}
\includegraphics[width=0.43\textwidth]{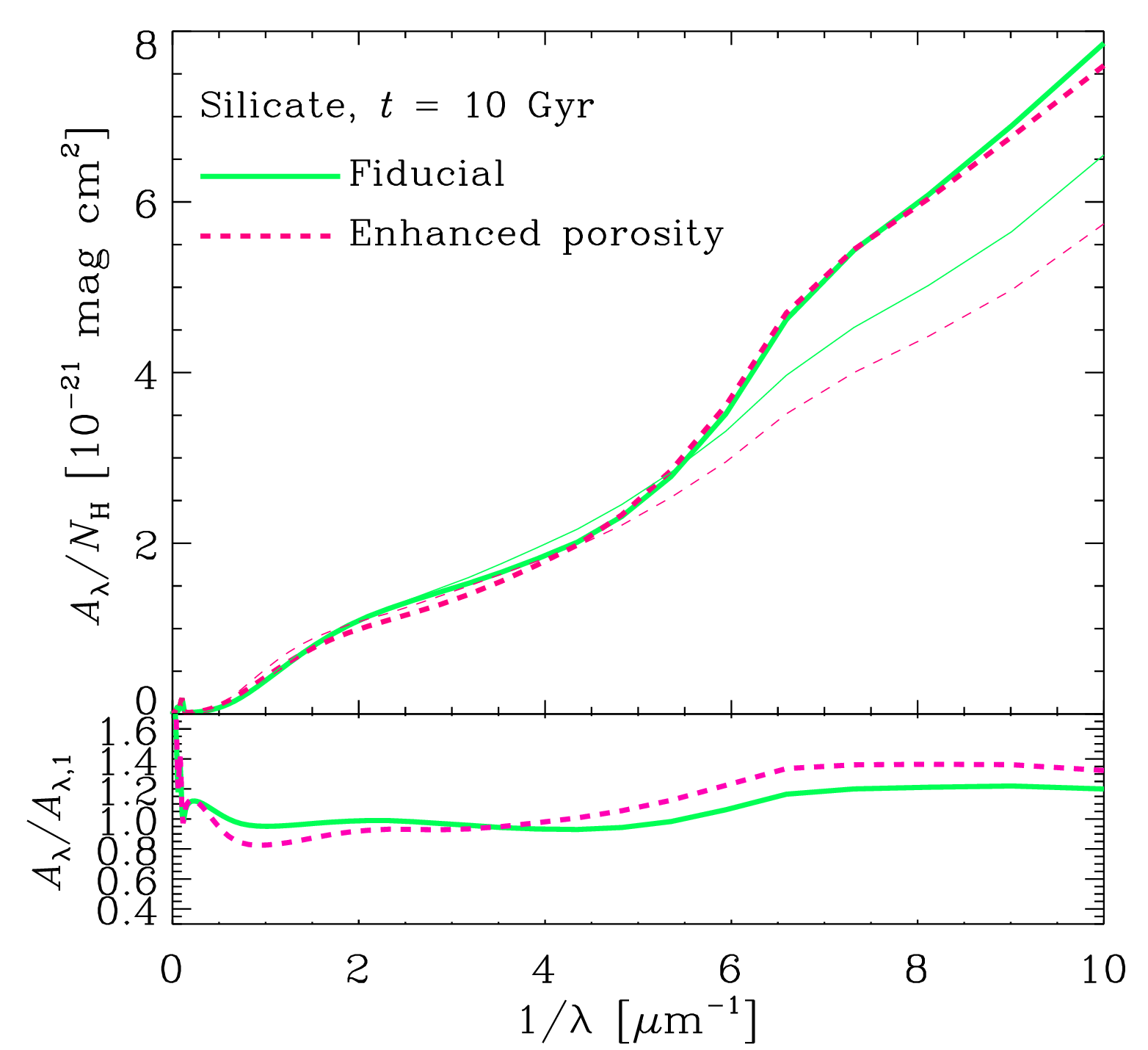}
\includegraphics[width=0.43\textwidth]{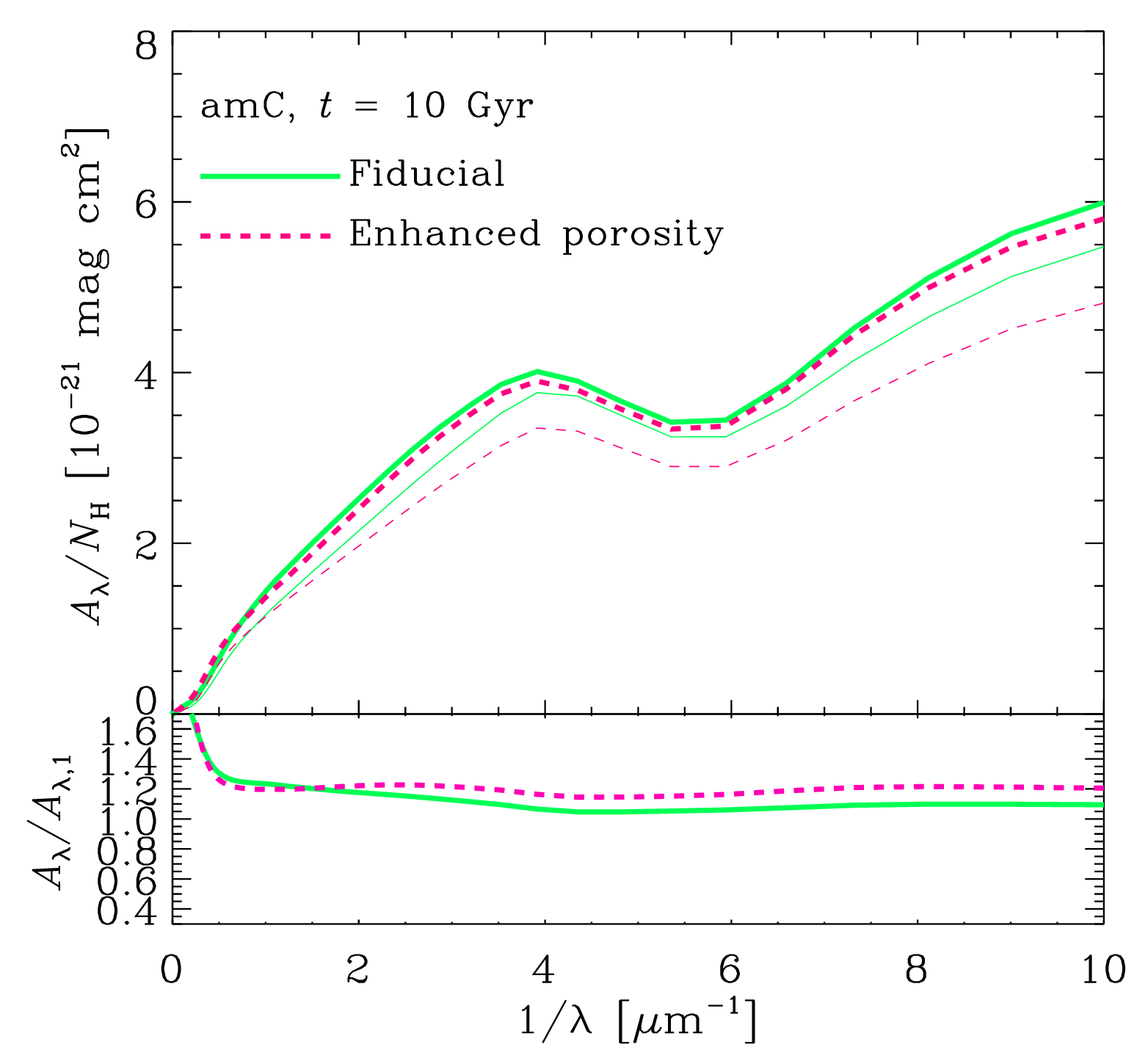}
\end{center}
\caption{Top: Grain size distributions at $t=10$ Gyr for the fiducial case and
the case with enhanced porosity (see text). The solid and dashed lines present
the results for the fiducial and the enhanced porosity cases, respectively.
Middle: The upper window displays the silicate extinction curves per hydrogen
($A_\lambda/N_\mathrm{H}$) corresponding to the grain size distributions and porosities
shown in the top with the same line species (the thick lines). The thin lines with the same line species
represent the extinction curves without porosity (i.e.\ $A_{\lambda ,1}$).
The lower window presents the ratio between the extinction curves
with and without porosity ($A_\lambda /A_{\lambda ,1}$).
The solid and dotted lines show the same models as in the upper window.
Bottom:
Same as the middle panel but for amC.
\label{fig:size_comp}}
\end{figure}

Next, we change $E_\mathrm{roll}$, which affects the porosity directly.
The rolling energy is regulated by the product of $\xi_\mathrm{crit}$ and $\gamma$.
As mentioned in Section \ref{subsubsec:coag}, $\gamma$ depends on the grain material and
its value is also uncertain. There is a possibility that $\gamma$ is larger than assumed in this paper.
Also, a larger value of $E_\mathrm{roll}\propto\gamma\xi_\mathrm{crit}$ is interesting to examine
how much the effect of porosity is enhanced within the uncertainties in the parameters.
As discussed in Section \ref{subsubsec:coag}, $\xi_\mathrm{crit}$ can be 3 times larger ($\sim 30$ \AA)
and $\gamma$ can also be 4 times larger ($\sim 100$ erg cm$^{-2}$).
Thus, we examine the case where $E_\mathrm{roll}$ is 12 times larger than in the above calculations
(this case with larger $E_\mathrm{roll}$ is referred to as \textit{the case with enhanced porosity}).
The fiducial parameters are used ($\eta_\mathrm{dense}=0.5$ and $\tau_\mathrm{SF}=5$~Gyr).
The difference in the porosity appears after $t\sim 1$ Gyr, when
coagulation starts to affect the grain size distribution significantly.
The difference is the largest at the latest epoch; thus, we compare the results at
$t=10$ Gyr in Fig.\ \ref{fig:size_comp}.

From Fig.\ \ref{fig:size_comp} (top), we observe that the porosity is indeed enhanced
at $a_m\sim 0.1~\micron$
in the case of enhanced porosity.
This is because compaction does not occur efficiently.
Compaction is still significant in the submicron regime because the grain velocities are
sufficiently large. The enhanced porosity increases the grain cross-sections, so that the
coagulation becomes more efficient. As a consequence, the upper cutoff of the grain radius
is slightly larger.
In this case, the porosity indeed increases (the filling factor drops down
to $\sim 0.2$ at $a_m\sim 0.1~\micron$).

In Fig.\ \ref{fig:size_comp} (middle and bottom), we compare 
the extinction curves between the case with enhanced porosity and the fiducial case.
Because the grain size distributions are different between the two cases,
the extinction curves without porosity (i.e.\ forcing $\phi_m$ to be unity, shown by the thin lines)
are different. However, the extinction curves which take
the porosity into account (shown by the thick lines) are similar between the two cases.
This is because the larger porosity tends to enhance the extinction curve slopes
(as shown in the plot
of $A_\lambda /A_{\lambda ,1}$).
This steepening effect almost cancels out the flattening by the increase of large grains.
Thus, different efficiencies of porosity formation lead to different grain size distributions and
porosities, but can predict very similar extinction curves. Because of the similar extinction
curves,
if we use the extinction curves predicted by the case of enhanced porosity
for the fitting to the MW extinction curve,
the resulting best fit solution and the
mass fraction of each component are almost the same as those shown in Section \ref{sec:MW}
(Table \ref{tab:fitting}; indicated by `Large porosity').
%%This is naturally expected because the
%%extinction curves are not altered by the enhanced porosity.
%%Therefore, within the parameter ranges adopted in this paper,
%%the porosity does not influence the fitting to the Milky Way extinction curve significantly.

%%The porosity effectively increases the sizes (characteristic radii) of grains, so that
%%it may enhance the grain--grain collision rate. Thus, we expect that the
%%grain size distribution is affected by the evolution of porosity. To examine how the porosity
%%influences the evolution of grain size distribution,
%%we compare models with and without enhanced porosity; for the latter, we
%%adopt large values for $\xi_\mathrm{crit}(=30~\text{\AA})$ and
%%$\gamma (=100~\mathrm{erg~cm^{-2}})$ as also done in the previous subsection.

In summary, the enhancement of grain radius by porosity affects the grain size
distribution through the increase of grain cross-sections, which leads to more large grains
through more efficient coagulation.
However, the effect on
the extinction curve is not obvious: Although the increase of large grains by porosity
tends to flatten the extinction curve, the porosity can steepen it. These two
effects can cancel out each other.
To distinguish the two effects, calculating dust emission SEDs
would be useful, particularly because
the porosity also affects the far-IR opacity \citep[e.g.][]{Voshchinnikov:2006aa,Ysard:2012aa,Ysard:2018aa}
while the mid-IR emission is sensitive to the grain size distribution
\citep[e.g.][]{Desert:1990aa,Dwek:1997aa,Draine:2001aa,Li:2001aa,Hirashita:2020ad}.
Starlight polarization may also be useful to constrain the porosity \citep{Draine:2021aa},
although non-sphericity is also essential in modelling polarization.
{\citet{Draine:2021aa} obtained low porosity ($\phi_m\lesssim 0.5$)
when a moderate axial ratio is assumed for spheroidal aligned grains. Considering that
their porosity constraint is applicable for $a_m\gtrsim 0.05~\micron$, where grain alignment is
efficient, our model predicts $\phi_m\lesssim 0.5$ in the fiducial model. Compaction plays an
important role in keeping the porosity small at large grain radii (Section \ref{subsec:gsd}).}
For further quantitative comparisons with observational data,
the dust emission SED and the starlight polarization are
worth modelling by extending our framework.

\subsection{Prospect for the fitting to the Milky Way extinction curve}\label{subsec:improve}

In Section \ref{sec:MW}, we showed that our fiducial model
with the additional inclusion of PAHs is broadly
successful in fitting the MW extinction curve. The shape of the best-fit extinction curve
does not significantly change even if we impose
$\phi_m=1$ (no porosity). This implies that our porosity model
does not significantly alter
the previous fitting to the MW extinction curve.
As shown in the previous subsection,
the uncertainties in the parameters that affect the porosity do not substantially
alter the output extinction curves.

We also note that
the 2175 \AA\ carriers are modelled by an \textit{ad hoc} PAH component, which
is not treated by the evolution model in this paper. \citet{Hirashita:2020aa} modelled
PAHs as small aromatic carbon grains in their evolution model of grain size distribution
\citep[see also][]{Seok:2014aa,Rau:2019aa}. However, PAHs cannot be
included in our model because their porosity is not well defined. Moreover, as mentioned
above, graphite, which is
another candidate of the 2175 \AA\ carriers, has a problem since its central wavelength shifts
as porosity increases. The difficulty in fitting the 2175 \AA\ carriers to our framework implies that
we need some special treatment for them; for example, the bulk material properties assumed in this
paper may not be applicable to them.
\citet{Hirashita:2020ad} also
showed that enhancement of the diffuse gas fraction is necessary to explain
the PAH emission but is not needed for small grain emission.
If the major formation sites of PAHs are segregated from those of other dust components,
our one-zone model is not capable of treating the PAH formation appropriately.
{Moreover, hydrodynamic effects such as enhanced densities caused
by supersonic turbulence could further promote interstellar processing as mentioned in
Section \ref{sec:model} \citep[e.g.][]{Hopkins:2016aa}.
This effect could not only enhance the spatial inhomogeneity in the grain size distribution
but also accelerate shattering, coagulation, and accretion, which may lead to
faster evolution of grain size distribution than predicted in this paper.}
In parallel, as done by \citet{Hirashita:2020ad}, it is also useful to predict
mid-IR emission SED, where PAHs have prominent features \citep[e.g.][]{Tielens:2008aa},
to further constrain the PAH abundance.

\section{Conclusions}\label{sec:conclusion}

We formulate and compute the evolution of grain size distribution and filling factor (porosity)
using a one-zone galaxy evolution model. We treat all the following processes considered previously to
calculate the evolution of grain size distribution: stellar dust production,
dust destruction by SN shocks sweeping the ISM, dust disruption by shattering in the diffuse ISM,
grain growth by coagulation in the dense ISM, and dust growth by the accretion of gas-phase metals
in the dense ISM. We extend the framework to include the evolution of grain porosity.
We assume that the dust grains formed in the stellar ejecta are compact.
For coagulation and shattering, we solve moment equations derived from
the 2-dimensional Smoluchowski equation by adopting the 
volume-averaging approximation; that is, the grain volume is represented by an averaged value
for each grain mass \citep{Okuzumi:2009aa,Okuzumi:2012aa}.
{In particular, coagulation is assumed to be the source of porosity.}
H21 already provided detailed results for shattering and
coagulation, and
we newly include the porosity evolution with accretion and SN destruction by
developing the moment equations based on the conservation of the grain number.
We assume that dust destruction does not change the porosity while
the newly condensed portion in accretion is compact.
As a consequence, our model is able to treat the grain size distribution and the porosity
{simultaneously}, and to predict the grain-size-dependent porosity in the entire evolutionary history
of a galaxy.

Since our model neglects the spatial variation within the galaxy,
we simply adopt a constant mass ratio between the diffuse and dense ISM by specifying
the dense gas fraction $\eta_\mathrm{dense}$ (the diffuse ISM occupies $1-\eta_\mathrm{dense}$).
We also vary the star formation time-scale $\tau_\mathrm{SF}$ to regulate the chemical
enrichment time-scale. We adopt $\eta_\mathrm{dense}=0.5$ and $\tau_\mathrm{SF}=5$ Gyr
for the fiducial values.

In the fiducial case, the result is described as follows. The evolution of grain size distribution
is similar to the case without porosity evolution as previously calculated by HA19.
In the early epoch, the grain size distribution is dominated by
large grains produced by stars and shattering gradually produces small grains
at $t\sim 0.3$~Gyr. In this phase, the grains are compact because these dominant processes
(stellar dust destruction and shattering) do not produce any porosity.
Porosity appreciably increases around
$t\sim 1$~Gyr when coagulation starts to become efficient owing to the increased abundance
of small grains by accretion. The filling factor decreases down to $\sim 0.3$
at $a_m\sim 0.03~\micron$, and reaches an equilibrium.
Compaction increases the filling factor at $a_m>0.03~\micron$.
The grain size distribution converges
to an MRN-like shape at $t\gtrsim 3$ Gyr.

The filling factor at $t\gtrsim 1$ Gyr is strongly affected by $\eta_\mathrm{dense}$.
The filling factor drops to 0.2 around $a_m\sim 0.03~\micron$ when $\eta_\mathrm{dense}$ is
low (0.2). In this case, shattering is more enhanced and coagulation is less efficient than in the fiducial
case. Although coagulation is the source of porosity, shattering produces small grains, from which
porous grains are built up by coagulation. This confirms the conclusion
in H21 that strong shattering with weak coagulation is favourable for creating grain porosity.
In the case of $\eta_\mathrm{dense}=0.8$,
large sub-micron grains are efficiently formed by coagulation, and the filling factor is relatively large
($\sim 0.5$ at $a_m\sim 0.03~\micron$).

We also examine the dependence on $\tau_\mathrm{SF}$. The filling factor achieved after the
evolution (e.g.\ at $t\gtrsim 3$ Gyr) is insensitive to $\tau_\mathrm{SF}$. We confirm that the
time-scale of interstellar processing scales as $\tau_\mathrm{SF}^{1/2}$ \citep{Hirashita:2020aa}.
That is,
a similar grain size distribution and porosity are obtained at the same $t/\sqrt{\tau_\mathrm{SF}}$.
This scaling is not exact (especially for the total dust abundance)
since the overall dust enrichment time-scale is scaled as
$\propto\tau_\mathrm{SF}$.

To investigate the effect of porosity evolution on observed dust properties,
we calculate optical--UV extinction curves using the effective medium theory for silicate and
amC separately.
Porosity enhances the far-UV extinction by $\sim$20 per cent and decreases the near-UV extinction
by $\sim$10 per cent for silicate in the fiducial case.
As a consequence, porosity makes the silicate extinction curve steeper.
The extinction of amC is enhanced by $\sim$10 per cent in the UV and $\sim 20$ per cent
in the optical, and the overall slope is not
sensitive to the porosity.
The decrease/increase of silicate extinction in the near/far-UV becomes larger
for $\eta_\mathrm{dense}=0.2$ and
smaller for $\eta_\mathrm{dense}=0.8$, reflecting the difference in the porosity achieved.
The wavelength range of diminished silicate extinction becomes wider for smaller $\eta_\mathrm{dense}$.

The above predictions will serve to examine or constrain the porosity evolution against the observed
extinction curve in the MW. The 2175~\AA\ bump is difficult to model in our framework,
since porous graphite grains show a shift in the peak wavelength of the bump, which is not observed in the
MW. Thus, we model the bump separately from the
two components (silicate and amC), using PAHs, and leave the origin of the bump carriers
for future work. We use the calculation results for the fiducial case at $t=10$ Gyr, and
fit the MW extinction curve. We find that the porosity does not affect the goodness of fitting.
This means that the grain size distribution, rather than the porosity, is more important in
fitting the MW extinction curve (as long as the porosity values calculated in our model are applicable).
Our fitting also requires similar amounts of silicate and amC, and the total dust-to-gas ratio
of $\sim 5\times 10^{-3}$. We also used the results for non-fiducial values
for $\eta_\mathrm{dense}$ (= 0.2 and 0.8) at $t=10$ Gyr to fit the MW extinction curve.
For $\eta_\mathrm{eta}=0.2$,
because the extinction curve of amC is steep enough, the overall slope of the MW extinction
curve is mostly reproduced by amC; thus, the fitting result predicts
negligible amount of silicate, which is not consistent with the interstellar depletion.
For $\eta_\mathrm{dense}=0.8$, the large-grain-dominated
size distribution predicts too flat an extinction curve in the optical and near-IR.
Overall, the
grain size distribution plays a more dominant role than the porosity in determining the
fitting solution for the MW extinction curve.

Grain porosity also affects the evolution of grain size distribution because it effectively increases the grain
cross-sections. To examine this effect, we increase
the values of the parameters ($\gamma$ and $\xi_\mathrm{crit}$)
that regulate the rolling energy. If the rolling energy is $\sim$ ten times higher (allowing for the
uncertainties in the parameters),
the abundance of large grains slightly increases.
This is because less compaction leads to
higher porosity at $a_m\gtrsim 0.1~\micron$, which increases the coagulation rate.
The resulting extinction curves are not necessarily flatter even with the larger
abundance of large grains. This is because larger porosity tends to make the
extinction curve steeper (especially for silicate), which counteracts the flattening of
extinction curve by the enhanced large-grain abundance.
Thus, although the enhancement of coagulation by porosity affects the grain size distribution,
it does not necessarily change the extinction curve significantly.

The calculations in this paper serve to give a basis for further extension of the model
to, for example, dust emission SEDs.
Furthermore, a modelling effort of including the 2175 \AA\ bump carriers is necessary.
If PAHs contribute not only to the 2175 \AA\ bump but also to mid-IR emission,
predictions of dust emission SEDs are a crucial step towards a comprehensive understanding
of dust evolution in galaxies. Polarization could also be important to predict, since it
may also constrain the porosity \citep{Draine:2021aa}.
We could also extend our predictions to high redshift galaxies to
investigate the evolution of dust porosity as well as grain size distribution
(see \citealt{Liu:2019aa} for a model at high redshift) in the
history of the Universe.

\section*{Acknowledgements}
 
We are grateful to the anonymous referee, L. Pagani, and B. T. Draine for useful comments.
HH thanks the Ministry of Science and Technology (MOST) for support through grant
MOST 107-2923-M-001-003-MY3 and MOST 108-2112-M-001-007-MY3, and the Academia Sinica
for Investigator Award AS-IA-109-M02.
VBI acknowledges the support from the RFBR grant 18-52-52006 and
the SUAI grant FSRF-2020-0004.

\section*{Data Availability}

Data related to this publication and its figures are available on request from the corresponding author.



%\bibliographystyle{mnras}
\bibliographystyle{mnras}
\bibliography{/Users/hirashita/bibdata/hirashita}
%%\bibliography{hirashita}

%%%%%%%%%%%%%%%%% APPENDICES %%%%%%%%%%%%%%%%%%%%%

\appendix

\section{Derivation of the equations for destruction and accretion}\label{app:derivation}

We derive equations (\ref{eq:dest_acc_m}) and (\ref{eq:dest_acc_V}) for dust destruction
by SN shocks and dust growth by accretion. The most important characteristic of these processes is
the conservation of the grain number in the $(m,\, V)$ space (except at the boundary)
\citep[see][but extended to the 2-dimensional space]{Hirashita:2011aa,Mattsson:2016aa}:
\begin{align}
\frac{\upartial f(m,\, V,\, t)}{\upartial t}+\frac{\upartial}{\upartial m}[\dot{m}f(m,\, V,\, t)]
+\frac{\upartial}{\upartial V}[\dot{V}f(m,\, V,\, t)]=0.\label{eq:cont}
\end{align}
Here the ranges of $m$ and $V$ are assumed to be $0\leq m<\infty$ and
$0\leq V<\infty$.
We take the zeroth moment of this equation (but multiplied by $m$) as
\begin{align}
\frac{\upartial}{\upartial t}\int mf\,\mathrm{d}V+\frac{\upartial}{\upartial m}\int\dot{m}mf\,\mathrm{d}V
-\int\dot{m}f\,\mathrm{d}V+\left[\dot{V}mf\right]_0^\infty =0,\label{eq:mom_zero}
\end{align}
where $[F(V)]_0^\infty\equiv F(\infty )-F(0)$, and we omit the variables in $f$
and the integration range $[0,\,\infty ]$.
The last term on the left-hand side in equation (\ref{eq:mom_zero}) vanishes since
$m=0$ at $V=0$ and $f$ is expected to approach to zero quickly enough as $V\to\infty$
(precisely speaking, we assume that
$\dot{V}$ is finite for $V\in [0,\,\infty]$ and that $mf\to 0$ as $V\to\infty$).
Next, we take the first moment of equation (\ref{eq:cont}) and obtain
\begin{align}
\frac{\upartial}{\upartial t}\int fV\,\mathrm{d}V+\frac{\upartial}{\upartial m}\int\dot{m}fV\,\mathrm{d}V
-\int\dot{V}f\,\mathrm{d}V+[V\dot{V}f]_0^\infty =0.\label{eq:mom_first}
\end{align}
The last term on the left-hand side of this equation vanishes for the same reasons as above.

We define the following mean quantity of $Q(m,\, V,\, t)$ ($Q$ is an arbitrary function of
$m$, $V$, and $t$):
\begin{align}
\bar{Q}(m,\, t)\equiv
\frac{{\displaystyle \int} Q(m,\, V,\, t)\, f((m,\, V,\, t)\,\mathrm{d}V}
{{\displaystyle \int} f(m,\, V,\, t)\,\mathrm{d}V}.\label{eq:mean}
\end{align}
Using $\tilde{n}(m,\, t)$ defined in equation (\ref{eq:n}) together with some mean quantities
(following equation \ref{eq:mean}),
we obtain from equations (\ref{eq:mom_zero}) and (\ref{eq:mom_first})
\begin{align}
\frac{\upartial}{\upartial t}[m\tilde{n}(m,\, t)]+\frac{\upartial}{\upartial m}[\bar{\dot{m}}(m,\, t)m\tilde{n}(m,\, t)]
-\bar{\dot{m}}(m,\, t)\tilde{n}(m,\, t)=0,\label{eq:mom0}
\end{align}
\begin{align}
\frac{\upartial}{\upartial t}[\bar{V}(m,\, t)\tilde{n}(m,\, t)]+
\frac{\upartial}{\upartial m}[\overline{\dot{m}V}(m,\, t)\tilde{n}(m,\, t)]
-\bar{\dot{V}}(m,\, t)\tilde{n}(m,\, t)=0.\label{eq:mom1}
\end{align}

To close this hierarchy of moment equations, we adopt the volume-averaging assumption
\citep{Okuzumi:2009aa};
that is, the volume is replaced with the mean value at each $m$.
This approximation is mathematically expressed as
$f(m,\, V)=\tilde{n}(m)\,\delta [V-\bar{V}(m)]$, where $\delta$ is Dirac's delta function.
Using this expression, we obtain
\begin{align}
\bar{\dot{m}}(m,\, t)=\dot{m}(m,\,\bar{V}(m,\, t),\, t),\label{eq:dotm}
\end{align}
\begin{align}
\overline{\dot{m}V}(m,\, t)=\dot{m}(m,\,\bar{V}(m,\, t),\, t)\bar{V}(m,\, t),
\end{align}
\begin{align}
\bar{\dot{V}}(m,\, t)=\dot{V}(m,\,\bar{V}(m,\,t),\, t).\label{eq:dotV}
\end{align}

Using equations (\ref{eq:dotm})--(\ref{eq:dotV}) together with equations (\ref{eq:rho_def})
and (\ref{eq:psi_def}) for the definitions of $\varrho$ and $\psi$, equations (\ref{eq:mom0})
and (\ref{eq:mom1}) are reduced to equations (\ref{eq:dest_acc_m}) and (\ref{eq:dest_acc_V}).
Note that in the main text, we simplify the notations as
$\dot{m}\equiv\dot{m}(m,\,\bar{V}(m,\, t),\, t)$ and
$\dot{V}\equiv\dot{V}(m,\,\bar{V}(m,\,t),\, t)$.

\section{Extinction curves for various parameter values}
\label{app:ext_dependence}

\begin{figure}
\includegraphics[width=0.48\textwidth]{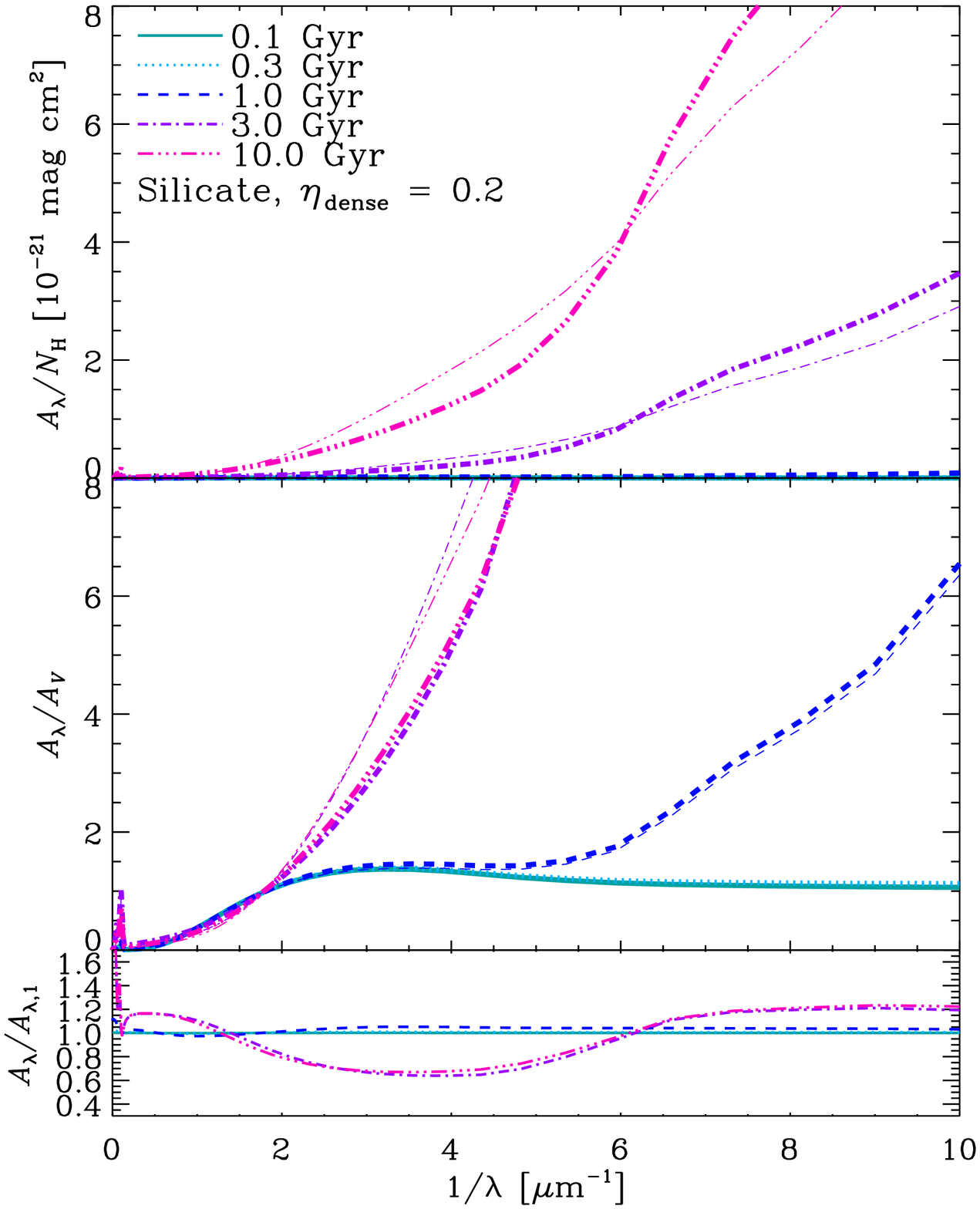}
\includegraphics[width=0.48\textwidth]{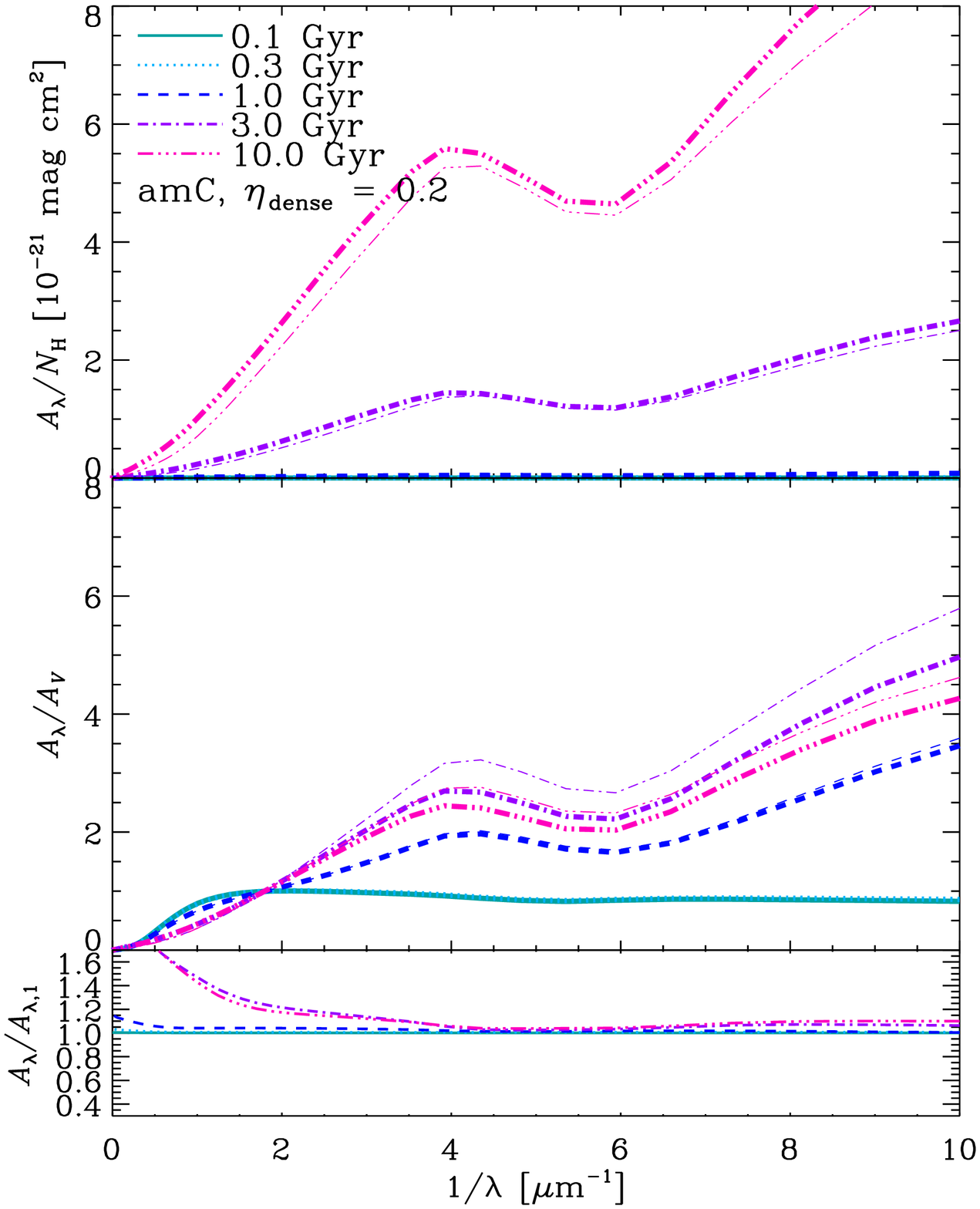}
\caption{Same as Fig.\ \ref{fig:ext} but for $\eta_\mathrm{dense}=0.2$.
\label{fig:ext_eta02}}
\end{figure}

\begin{figure}
\includegraphics[width=0.48\textwidth]{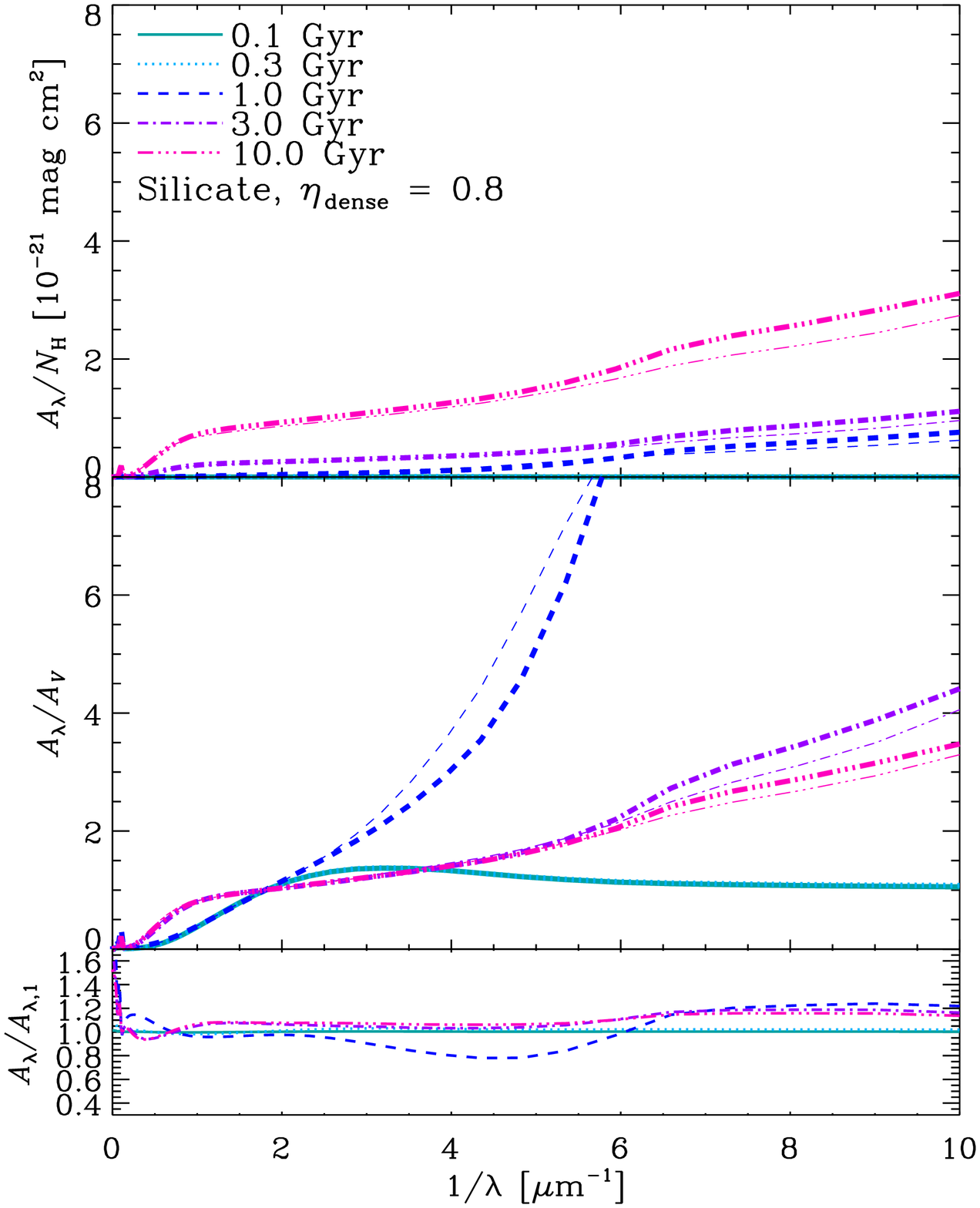}
\includegraphics[width=0.48\textwidth]{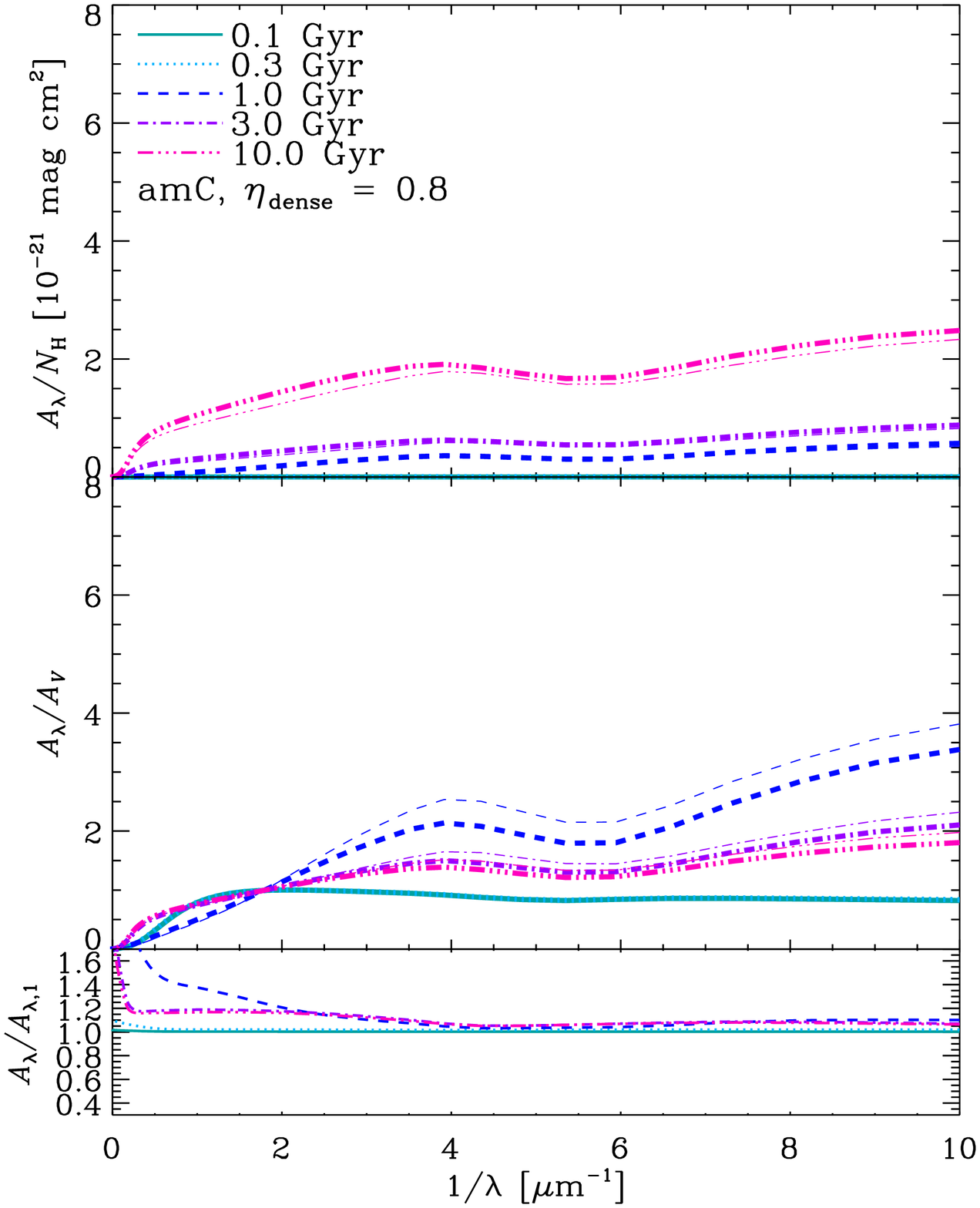}
\caption{Same as Fig.\ \ref{fig:ext} but for $\eta_\mathrm{dense}=0.8$.
\label{fig:ext_eta08}}
\end{figure}

\begin{figure}
\includegraphics[width=0.48\textwidth]{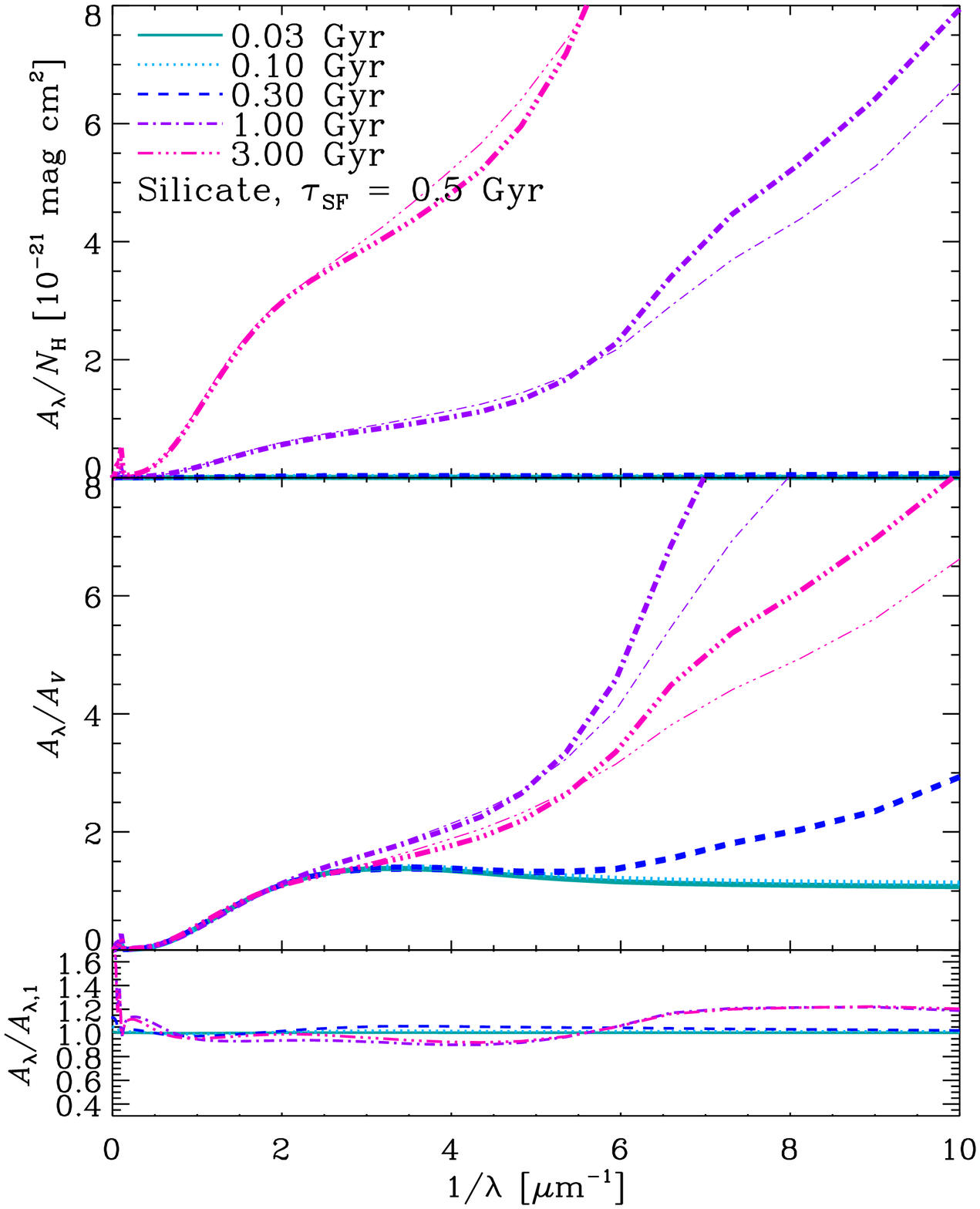}
\includegraphics[width=0.48\textwidth]{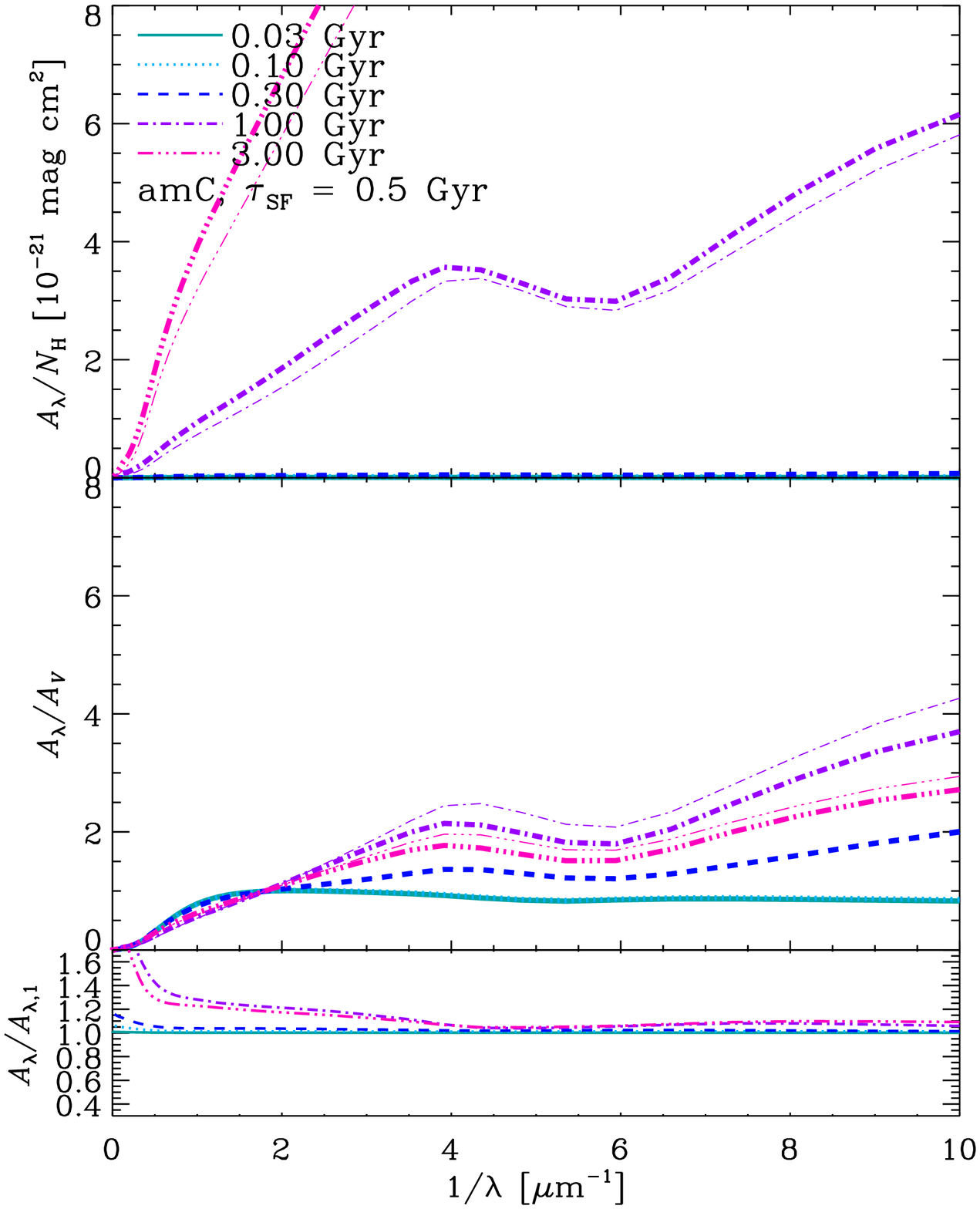}
\caption{Same as Fig.\ \ref{fig:ext} but for $\tau_\mathrm{SF}=0.5$ Gyr.
\label{fig:ext_tau05}}
\end{figure}

\begin{figure}
\includegraphics[width=0.48\textwidth]{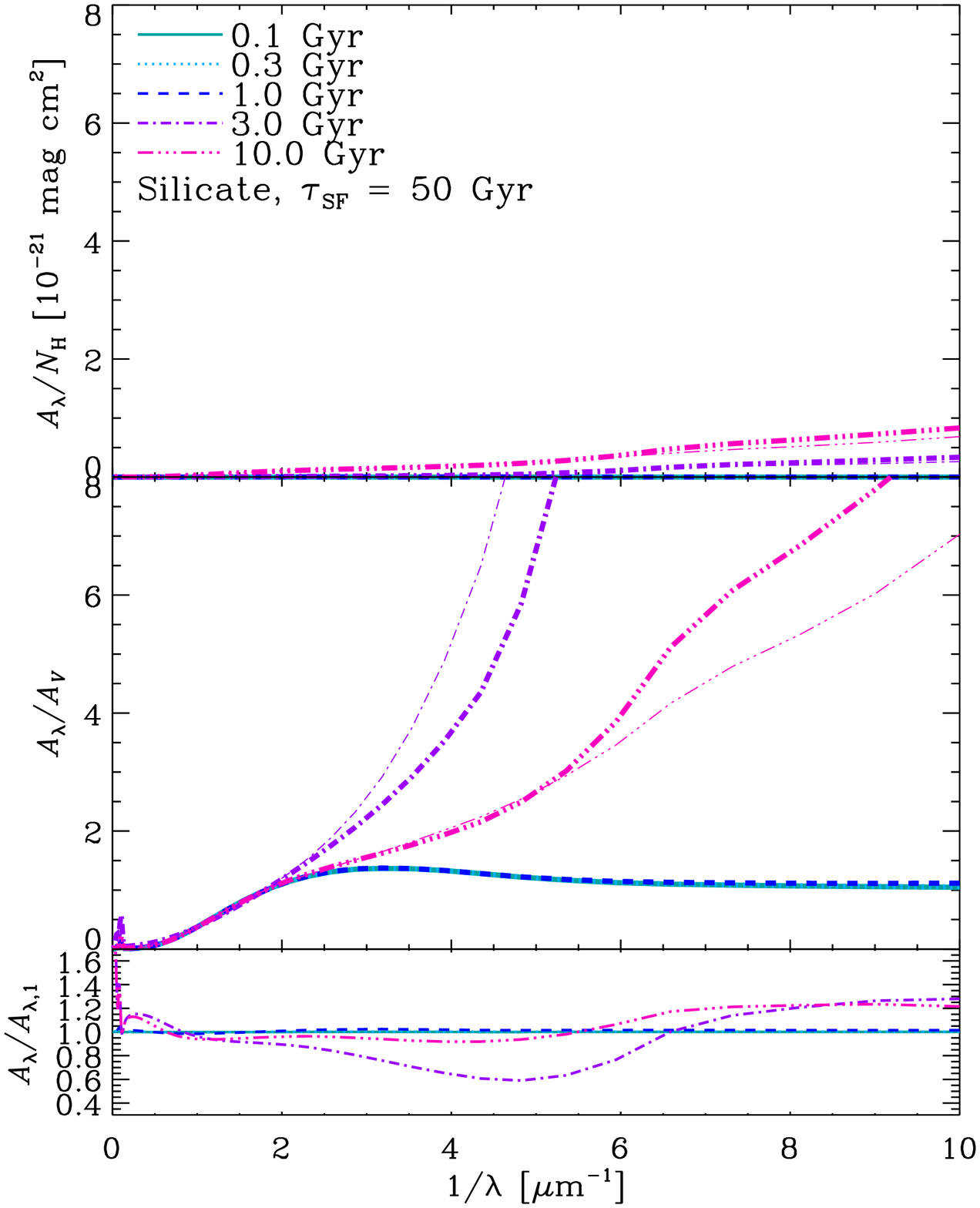}
\includegraphics[width=0.48\textwidth]{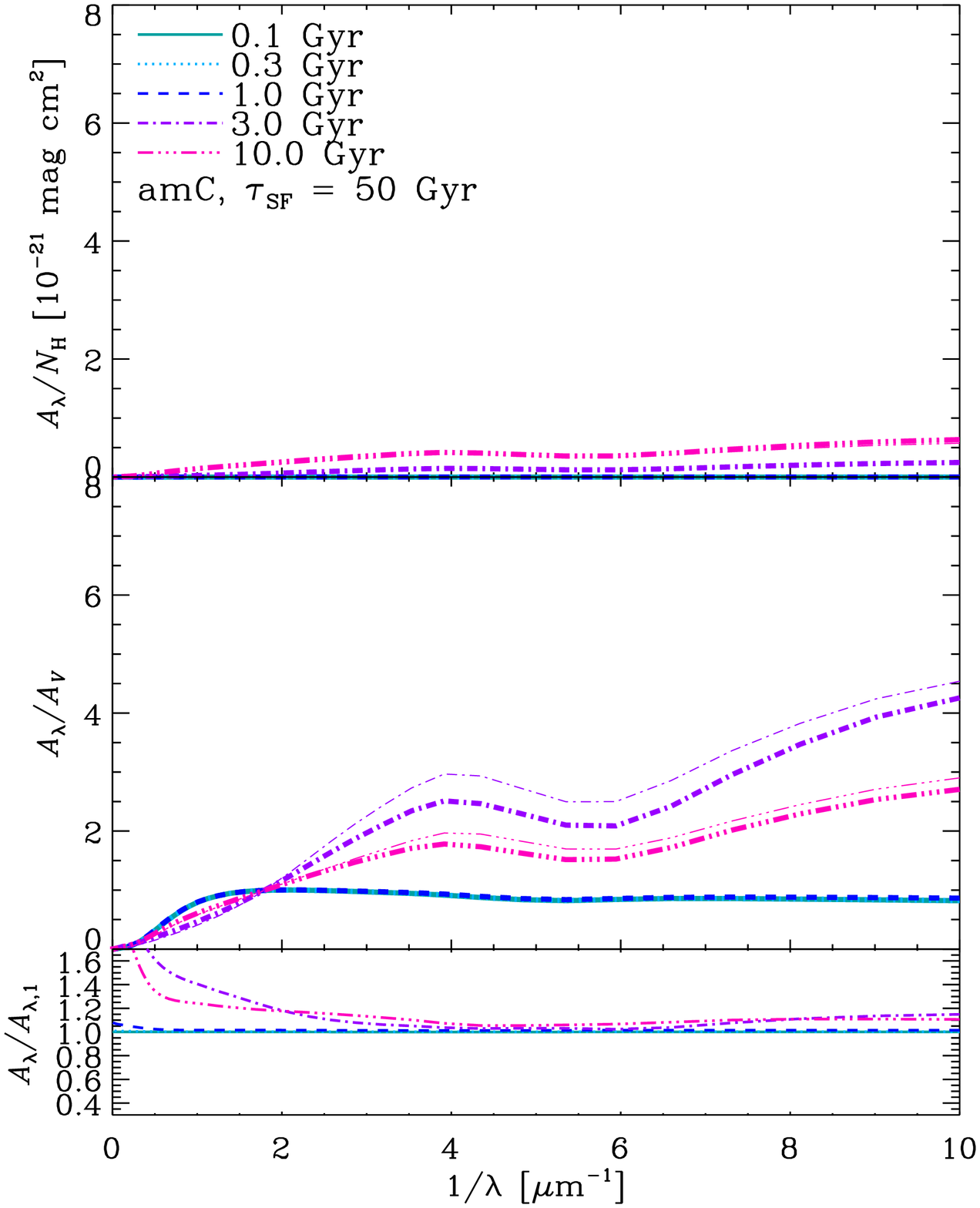}
\caption{Same as Fig.\ \ref{fig:ext} but for $\tau_\mathrm{SF}=50$ Gyr.
\label{fig:ext_tau50}}
\end{figure}

We show the extinction curves for silicate and amC together with the
increment of extinction by porosity ($A_\lambda /A_{\lambda ,1}$) for the cases other
than the fiducial case ($\eta_\mathrm{dense}=0.5$ and $\tau_\mathrm{SF}=5$ Gyr),
which is shown in the text (Section \ref{subsec:ext_result};
Fig.\ \ref{fig:ext}). We present the cases with $\eta_\mathrm{dense}=0.2$ and 0.8
(with $\tau_\mathrm{SF}=5$ Gyr) in Figs.\ \ref{fig:ext_eta02} and \ref{fig:ext_eta08},
respectively. We also show the dependence on $\tau_\mathrm{SF}$ with a fixed
$\eta_\mathrm{dense}(=0.5)$ ($\tau_\mathrm{SF}=0.5$ and 50 Gyr
in Figs.\ \ref{fig:ext_tau05} and \ref{fig:ext_tau50}, respectively).
The discussions on these figures
are provided in Section \ref{subsec:ext_result}.

% Don't change these lines
\bsp	% typesetting comment
\label{lastpage}
\end{document}